\documentclass[a4paper,12pt]{article}
\usepackage{latexsym}
\usepackage{amsmath}
\usepackage{epsfig}
\usepackage{cite}
\usepackage{subfigure}
\begin{document}



\input paperdef


\begin{flushright}
DCPT/09/154\\
IPPP/09/77\\
\end{flushright}

\vspace{1em}

\begin{center}

{\large\sc {\bf Precise Predictions for Higgs Production
in Neutralino Decays}}

\vspace*{0.3cm}

{\large\sc {\bf in the Complex MSSM}}

\vspace{1.0cm}

{\sc A.C.~Fowler%
\footnote{
email: A.C.Fowler@durham.ac.uk
}%
~and G.~Weiglein%
\footnote{
email: Georg.Weiglein@durham.ac.uk
}%
}

\vspace*{0.7cm}

IPPP, University of Durham, Durham DH1~3LE, UK

\end{center}

\vspace*{0.1cm}

\begin{abstract}
\noindent
Complete one-loop results are obtained for the class of processes 
$\tilde{\chi}^0_i\rightarrow \tilde{\chi}^0_j h_a$ in the MSSM where all
parameters entering this process beyond lowest order are allowed to have
arbitrary $\cp$-violating phases. An on-shell renormalisation scheme is worked out for the
chargino--neutralino sector that properly takes account of imaginary
parts arising from complex parameters and from absorptive parts of loop
integrals.
The genuine vertex contributions to the neutralino decay amplitudes
are combined with two-loop propagator-type corrections for the 
outgoing Higgs boson.
In this way the currently most precise prediction for this class of
processes is obtained. The numerical impact of the genuine vertex
corrections is studied for several examples of $\cp$-conserving and 
$\cp$-violating scenarios. We find that significant effects on the decay widths
and branching ratios are possible even in the $\cp$-conserving MSSM.
In the $\cp$-violating CPX benchmark scenario the corrections to the
decay width are found to be particularly large, namely, of order $45\%$ for a Higgs mass of $40\,\mathrm{GeV}$.
This parameter
region of the CPX scenario where a very light Higgs boson is unexcluded
by present data is analysed in detail. 
We find that in this parameter region, which will be difficult to cover by
standard Higgs search channels at the LHC, the branching ratio for the
decay $\tilde{\chi}^0_2\rightarrow \tilde{\chi}^0_1 h_1$ is large. This
may offer good prospects to detect such a light Higgs boson in cascade
decays of supersymmetric particles.
\end{abstract}

\def\thefootnote{\arabic{footnote}}
\setcounter{footnote}{0}

\newpage


\section{Introduction}

One of the main goals of physics at the Large Hadron Collider (LHC) will
be to shed light on the mechanism of electroweak symmetry breaking (EWSB).
The most popular realisation of EWSB in theoretical models is the Higgs
mechanism, which gives rise to at least one fundamental scalar particle in
the spectrum. While in the Standard Model (SM), Higgs physics is
determined by a single parameter, the mass of the Higgs boson,
$\MH$, a much richer phenomenology is possible in extensions of the SM.

In the Minimal Supersymmetric Standard Model (MSSM), the Higgs
sector is characterised at lowest order by two new parameters instead of
one.  
The spectrum contains five physical
Higgs bosons, the properties of which may differ significantly from those of a SM Higgs. At lowest order the neutral Higgs
bosons of the MSSM are $\cp$-eigenstates, so that there are two
$\cp$-even Higgs bosons, $h$ and $H$, a $\cp$-odd Higgs boson, $A$, and
two charged Higgs bosons, $H^\pm$. Higher-order contributions in the
MSSM Higgs sector yield large corrections to the masses
and couplings, and can also induce $\cp$-violation, so that
mixing can occur between $h,H$ and $A$ in the general case of complex
SUSY-breaking
parameters. If the mixing between the three neutral mass
eigenstates, denoted $h_1$, $h_2$ and $h_3$, is such that the coupling of the lightest Higgs boson, $h_1$, to
gauge bosons is significantly suppressed, this state can be very light
without being in conflict with the exclusion bounds from the Higgs
searches at LEP~\cite{Barate:2003sz,Schael:2006cr} and the 
Tevatron~\cite{Amsler:2008zzb}. In particular, in the CPX
benchmark scenario~\cite{Carena:2000ks} an unexcluded region remains in which 
$\MHe \approx 45 \gev$ and $\tan\beta \approx 7$~\cite{Schael:2006cr} (see also
\citere{Williams:2007dc} for a recent reevaluation with improved
theoretical predictions). This unexcluded parameter region with a
very light Higgs boson will also be difficult to cover at the LHC 
with the standard search channels~\cite{Buescher:2005re,Schumacher:2004da,Accomando:2006ga}.

While on the one hand a supersymmetric (SUSY) 
scenario such as the CPX scenario may have much worse prospects compared
to the SM case for 
Higgs searches at the LHC in the standard channels,
on the other hand additional Higgs production channels involving
SUSY particles may occur in such a case. In cascade decays of
heavier SUSY particles down to the lightest supersymmetric
particle (LSP), Higgs bosons can in particular be produced in decays of 
neutralinos and charginos,
via $\tilde{\chi}^0_i\rightarrow\tilde{\chi}^0_j
\,h,H\,\mathrm{or}\,A$ and $\tilde{\chi}^{\pm}_i\rightarrow\tilde{\chi}^0_j
H^{\pm}$, see e.g.\ \citeres{Datta:2003iz,Ball:2007zza}
for studies of these channels at the LHC in the MSSM with real
parameters.  These channels have also attracted recent interest for studies 
of scenarios with non-universal gaugino masses \cite{Bandyopadhyay:2008fp,Huitu:2008sa,Bandyopadhyay:2008sd}.
In the parameter regions of the CMSSM (the constrained
MSSM) and the NUHM (a generalisation of the CMSSM with a non-universal
Higgs mass parameter) which are currently favoured by electroweak precision 
data, $B$-physics observables and cosmological data,
an early discovery of the light Higgs boson from a neutralino decay
in a SUSY cascade could be
possible~\cite{Buchmueller:2008qe}.  
A related process to the production of a Higgs boson in the decay of a
neutralino is the decay of a heavy Higgs boson into two neutralinos, 
$H,A\rightarrow \tilde{\chi}^0_i \tilde{\chi}^0_j$.
This process, with a possible signature of four leptons plus missing
energy, can also be
phenomenologically important~\cite{Moortgat:2001pp,Bisset:2007mi}.
Concerning theoretical predictions for this class of processes, 
partial one-loop results have been published previously for the decays 
$H,A\rightarrow \tilde{\chi}^0_i\tilde{\chi}^0_j$ in both 
the Feynman-diagrammatic~\cite{Eberl:2001vb,Zhang:2002fu} and effective 
potential~\cite{Ibrahim:2008rq} approaches. These predictions did not 
include the full MSSM,
and the Feynman-diagrammatic calculations were restricted to the 
case of real parameters.

In the present paper we obtain predictions
for decays of a heavier neutralino into
a lighter neutralino and a neutral Higgs boson in the MSSM with complex
parameters, i.e.\ we consider the 
class of processes $\tilde{\chi}^0_i\rightarrow \tilde{\chi}^0_j h_a$,
where $h_a=h_1,h_2,h_3$.  Our calculations are also applicable to the related 
class of processes $h_a \rightarrow\tilde{\chi}^0_i\tilde{\chi}^0_j$.
Since higher-order contributions in the MSSM Higgs sector are known to 
be large, a proper inclusion of Higgs-sector corrections is
indispensable for a reliable prediction of this class of processes. 
The process-independent
corrections to the mass of the outgoing Higgs boson and
to the Higgs wave function normalisation can be
incorporated via an effective Born-type prediction for the neutralino
decay process, see \citeres{Frank:2006yh,Williams:2007dc,Hahn:2007ut}.
The genuine (process-specific) vertex corrections can also be very 
important.  This has recently been demonstrated in 
\citere{Williams:2007dc} for Higgs
cascade decay processes, $h_a \to h_b h_c$, in the CPX scenario, where 
the genuine vertex corrections were found to give rise to drastic
changes in the decay widths compared to the effective Born-type
predictions. In the neutralino decay processes, 
comprising just one instead of three external Higgs bosons, the genuine
vertex corrections are not expected to be quite as large as for the
Higgs cascade decays, but their effects can nevertheless be expected to
be non-negligible.

We use the Feynman-diagrammatic approach to evaluate
higher-order contributions to the processes 
$\tilde{\chi}^0_i\rightarrow \tilde{\chi}^0_j h_a$. Specifically,
we compute the
vertex corrections at the one-loop level, taking into account the 
contributions from all MSSM particles, and we combine these results with
state-of-the-art two-loop propagator-type corrections as implemented in
the code \texttt{FeynHiggs}~\cite{feynhiggs,fhrandproc,Frank:2006yh,mhcpv2l}.
In this way the currently most precise prediction for this class of
processes is obtained. We focus our treatment of $\cp$-violating phases
on those that are most relevant for Higgs
phenomenology, namely the phases of the trilinear couplings of the third
generation, $\phi_{A_{\mathrm{t,b},\tau}}$, and the gluino phase,
$\phi_{M_3}$ (these are also the phases chosen to be non-zero in the CPX
benchmark scenario~\cite{Carena:2000ks}; the gluino phase enters the 
predictions for the neutralino decays via two-loop Higgs propagator-type
contributions).
We use an on-shell scheme for the renormalisation in the chargino--neutralino sector. In the MSSM with 
complex parameters care has to be taken 
in the treatment of absorptive parts of loop integrals and 
imaginary parts of MSSM parameters, since 
products of such contributions can enter predictions for physical
observables already at the one-loop level. We have worked out a scheme
for the renormalisation in the chargino--neutralino sector where in- and
outgoing fermions receive different field renormalisation constants.

In our numerical discussion we concentrate in particular on the
parameter region in the CPX benchmark scenario where a light Higgs boson
is unexcluded by current data
(see also 
\citeres{Akeroyd:2003jp,Ghosh:2004cc,Bandyopadhyay:2007cp} for discussions
of other possible LHC search channels to access this parameter region), but 
we also give examples for the $\cp$-conserving case. Based
on our results, we investigate the phenomenology of Higgs
searches at the LHC in the channels 
$\tilde{\chi}^0_i\rightarrow \tilde{\chi}^0_j h_a$.
We briefly discuss the prospects for 
covering the unexcluded parameter region of the CPX scenario 
in this way.
\section{Lowest-order Result, Notations and Conventions}
\label{sectiontree}
We first lay out our notation for the Higgs and chargino--neutralino 
sectors of the MSSM, and use this to write down a formula for the tree-level 
decay width for $\tilde{\chi}^0_i\rightarrow \tilde{\chi}^0_j h^0_k$, where 
$h^0_k$ is one of the neutral MSSM Higgs bosons, $h$, $H$ or $A$.  
We also include notation for the sfermion sector which enters the
process at the one-loop level.
\subsection{Higgs Sector}
In the Higgs sector we follow the conventions of 
\citeres{Frank:2006yh,Williams:2007dc}.
We write the two Higgs doublets at tree level as
\begin{eqnarray}
\mathcal{H}_1&=\left( \begin{array}{cc}
                            v_1+\frac{1}{\sqrt{2}}(\phi_1-i \chi_1)\\
                            -\phi_1^{-}
                            \end{array} \right),\;\;
\mathcal{H}_2&=e^{i\xi}\left( \begin{array}{cc}
                            \phi_2^{+}\\
                            v_2+\frac{1}{\sqrt{2}}(\phi_2+i \chi_2)
                            \end{array} \right).
\end{eqnarray}
The tree-level physical states $h,H,A,H^{\pm}$ and unphysical Goldstone states 
$G,G^{\pm}$ are obtained from rotations by the mixing angles 
$\alpha$, $\beta_n$ and $\beta_c$ as shown,
\begin{equation}
\left( \begin{array}{c}
h \\
H \\
A \\
G \end{array} \right) =
\left( \begin{array}{cccc}
-\sin{\alpha} & \cos{\alpha} & 0 & 0 \\
\cos{\alpha}   & \sin{\alpha} & 0 & 0 \\
0 & 0 & -\sin{\beta_n} & \cos{\beta_n} \\
0 & 0 & \cos{\beta_n} & \sin{\beta_n} \end{array} \right) 
\left( \begin{array}{c}
\phi_1 \\
\phi_2 \\
\chi_1 \\
\chi_2 \end{array} \right) , 
\end{equation}
\begin{equation}
\left( \begin{array}{c}
H^{\pm} \\
G^{\pm} \end{array} \right)=
\left( \begin{array}{cc}
-\sin{\beta_c} & \cos{\beta_c} \\
\cos{\beta_c} & \sin{\beta_c}
\end{array} \right)
\left( \begin{array}{c}
\phi_1^{\pm} \\
\phi_2^{\pm}
\end{array} \right).
\end{equation}
As indicated by the null entries in the $4 \times 4$ mixing matrix above,
at tree level there is no $\cp$-violating mixing between the neutral 
Higgs bosons. Minimization of the Higgs potential and the requirement of 
vanishing tadpoles at tree level renders the phase $\xi=0$ and 
$\beta_n=\beta_c=\beta$, where $\tan\beta \equiv v_2/v_1$ is the ratio of 
the Higgs vacuum expectation values. The Higgs sector is characterised
by two input parameters (besides the gauge couplings), conveniently
chosen as $\tb$ and one of the Higgs-boson masses. For the latter,
the most convenient choice
in the case where the SUSY-breaking parameters are allowed to be complex 
is the mass of the charged Higgs boson, $\MHp$, since the three neutral 
Higgs bosons mix with each other once higher-order corrections are taken into
account.
\subsection{Chargino and Neutralino Sector}
At tree level, the physical chargino states, $\tilde{\chi}^\pm_i$
($i=1,2$), are Dirac spinors constructed from the mass eigenstates of
the $2 \times 2$ complex mass matrix $X$, which reads, 
in the wino-higgsino basis,
\begin{equation}
X=
\left( \begin{array}{cc}
M_2 & \sqrt{2} M_W\sin\beta  \\
\sqrt{2} M_W\cos\beta  & \mu
\end{array} \right) ,
\end{equation}
where $M_2$ and $\mu$ are the wino and higgsino mass parameters,
respectively.  The off-diagonal elements depend on parameters from other
sectors, namely $\tan\beta$ and $M_W$, the mass of the W boson.  The
mass matrix is diagonalised by two $2 \times 2$ complex 
unitary matrices $U$ and $V$,
where $U^{\ast}X
V^{\dagger}=\mathrm{diag}(m_{\tilde{\chi}^\pm_1},m_{\tilde{\chi}^\pm_2})$.  
Similarly, the neutralinos $\tilde{\chi}^0_i$, ($i=1,2,3,4$) are
Majorana spinors constructed from mass eigenstates of the $4 \times 4$ 
complex mass matrix $Y$, which reads, 
in the 
($\widetilde{B},\widetilde{W}^3,\widetilde{H}^{0}_1,\widetilde{H}^{0}_2$) 
basis:
\begin{equation}
Y=
\left( \begin{array}{cccc}
M_1 & 0 & -M_Z c_\beta s_W & M_Z s_\beta s_W \\
0   & M_2 & M_Z c_\beta c_W & -M_Z s_\beta c_W \\
-M_Z c_\beta s_W & M_Z c_\beta c_W & 0 & -\mu \\
M_Z s_\beta s_W & -M_Z s_\beta c_W & -\mu & 0 \end{array} \right) ,
\end{equation}
where $M_1$ is the bino mass parameter, $M_Z$ is the mass of the Z boson and 
$s_W \equiv \sin\theta_{W}$ is the sine of the weak mixing angle.  
We adopt the abbreviations $c_{\beta} \equiv \cos\beta$ and 
$s_{\beta} \equiv \sin\beta$.  Due to the Majorana nature of
neutralinos, only one $4 \times 4$ complex unitary matrix $N$ 
is required to diagonalise $Y$, where 
$N^{\ast}YN^{\dagger}=\mathrm{diag}(m_{\tilde{\chi}^0_1},m_{\tilde{\chi}^0_2},m_{\tilde{\chi}^0_3},m_{\tilde{\chi}^0_4})$.  
Besides parameters from other sectors, the masses and mixings of neutralinos and charginos can thus be described by three independent input parameters, $M_1$, $M_2$ and $\mu$.  If all three parameters are real, then $X$ and $Y$ can also be chosen to be real, while each of the rows of $N$ can be chosen to be purely real or purely imaginary such that all neutralino masses are positive.
\subsection{Sfermion Sector}
At tree level, the physical squark and charged slepton states,
$\tilde{f}_1$, $\tilde{f}_2$, are the mass eigenstates of a $2 \times 2$ complex mass matrix, which reads in the ($\tilde{f}_L$, $\tilde{f}_R$) basis for each flavour,
\begin{equation}
M_{\tilde{f}}=
\left( \begin{array}{cc}
M_L^2+m_f^2+M_Z^2\cos{2\beta} (I^f_3-Q_f s_W^2) & m_f X^{\ast}_f  \\
m_f X_f  & M_{\tilde{f}_R}^2+m_f^2+M_Z^2\cos{2\beta} Q_f s_W^2
\end{array} \right) ,
\label{sfermion}
\end{equation}
with
\begin{equation}
\label{Xf}
 X_f=A_f-\mu^{\ast} \left\{\cot\beta,\tan\beta\right\} ,
\end{equation}
where $\left\{\cot\beta,\tan\beta\right\}$ applies for up- and down-type
sfermions, respectively.  The soft SUSY-breaking parameters introduced in 
the sfermion sector are $M_L^2$ and $M_{\tilde{f}_R}^2$, which are real, and the trilinear coupling $A_f$, which can be complex.  
The phase $\phi_{A_f}$ can play an important role in loops involving 
the supersymmetric partners of the heavy third-generation SM fermions, $t,b,\tau$, where the term $m_f X_f$ appears in couplings of sfermions to Higgs bosons.  
\subsection{Tree-level Decay Width}
For the interaction of neutralinos with neutral Higgs bosons, the relevant piece of the Lagrangian can be written in terms of tree-level mass eigenstates as,
\begin{equation}
 \mathcal{L}= \frac{i}{2}\, h^0_k\, \overline{\tilde{\chi}^0_i}\,
[\omega_R C^R_{ijh^0_k}  +\omega_L (-1)^{\delta_{k3}} (-1)^{\delta_{k4}}
C^{L}_{ijh^0_k})]\, \tilde{\chi}^0_j ,
\label{lagrangian}
\end{equation}
where $\omega_{R/L}=\frac{1}{2}(1\pm\gamma_5)$, and $k$ labels neutral
Higgs bosons, i.e. $h^0_k=\{h,H,A,G\}$.  A minus sign appears between
the $\omega_R$ and $\omega_L$ terms for the $\cp$-odd Higgs states.  The couplings, $C^{R/L}_{ijh^0_k}$, are given by
\begin{eqnarray}
 C^R_{ijh^0_k}=C^{L^{\ast}}_{ijh^0_k}=\frac{e}{2 c_W s_W} c_{ijh^0_k} ,
\end{eqnarray}
where
\begin{eqnarray}
c_{ijh^0_k} &=&[(a_k N_{i3} + b_k N_{i4})(s_W N_{j1} - c_W N_{j2})+(a_k N_{j3}+b_k N_{j4})(s_W N_{i1} - c_W N_{i2})]\nonumber\\
a_k&=&\{-s_{\alpha},c_{\alpha},i s_{\beta_n},-i c_{\beta_n}\}\nonumber\\
b_k&=&\{-c_{\alpha},-s_{\alpha},-i c_{\beta_n},-i s_{\beta_n}\}.
\end{eqnarray}
The tree-level decay width $\Gamma^{\mathrm{tree}}$ for the two-body decay 
$\tilde{\chi}^0_i\rightarrow \tilde{\chi}^0_j h^0_k$, where $h^0_k=\{h,H,A\}$, can then be written as
\begin{equation}
\Gamma^{\mathrm{tree}}=\frac{1}{16 \pi
m_{\tilde{\chi}^0_i}^3}|C^R_{ijh^0_k}|^2\,\kappa
(m_{\tilde{\chi}^0_i}^2,m_{\tilde{\chi}^0_j}^2,m_{h^0_k}^2)\,[m_{\tilde{\chi}^0_i}^2+m_{\tilde{\chi}^0_j}^2-m_{h^0_k}^2+2
(-1)^{\delta_{k3}} m_{\tilde{\chi}^0_i} m_{\tilde{\chi}^0_j}] ,
\label{eqntree}
\end{equation}
with
\begin{equation}
 \kappa (x,y,z)=((x^2-y^2-z^2)^2-4 y z)^{1/2}.
\end{equation}
In order to obtain a prediction for the decay width at one-loop level,
the parameters appearing in the lowest-order result and the fields of
$\tilde{\chi}^0_i,\tilde{\chi}^0_j,h^0_k$ need to be renormalised.  We describe their renormalisation in the next section.
Note that the mixing matrix elements involving $\alpha$, $\beta_n$ and $N_{ij}$ are not renormalised in our scheme, and $\beta_n$ is set equal to $\beta$ only after the renormalisation has been carried out.
\section{One-loop Calculation for 
$\tilde{\chi}^0_i\rightarrow \tilde{\chi}^0_j h_a$ and Combination with
Higher-order Contributions}
\label{sectionloop}
\begin{figure}[htb!]
\centering
\includegraphics[height=3.5cm]{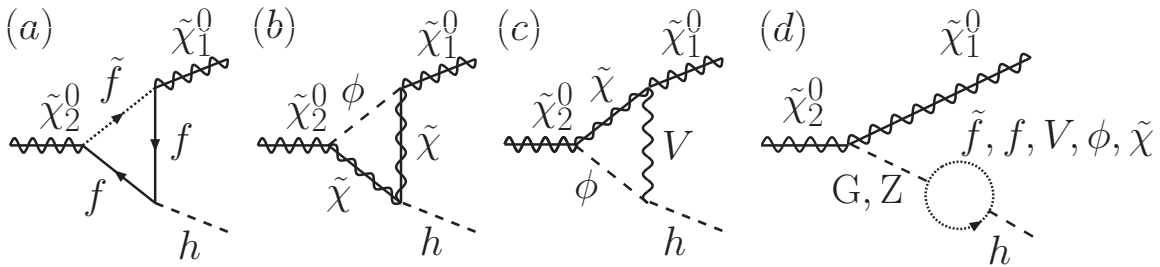}
\caption{\footnotesize{(a) Example of a 1PI vertex diagram for $\tilde{\chi}^0_2\rightarrow \tilde{\chi}^0_1 h$ involving fermions and sfermions.  There are UV-finite subsets of diagrams with the same generations and/or flavours; (b,c) Examples of 1PI vertex diagrams involving gauge bosons, Higgs bosons and their superpartners; (d) Examples of reducible G-Z 
mixing self-energy diagrams.  The particles are labelled according to
$f=q,l,\nu$, $\tilde{\chi}=\tilde{\chi}^0_i,\tilde{\chi}^{\pm}_i$, $\phi=h,H,A,G,H^{\pm},G^{\pm}$ and $V=\gamma,Z,W^{\pm}.$}}
\label{diag}
\end{figure}

We have calculated the full one-loop vertex corrections to the process 
$\tilde{\chi}^0_i\rightarrow \tilde{\chi}^0_j h_a$, where 
$h_a=\{h_1,h_2,h_3\}$, taking into account all sectors of the MSSM and
the full phase dependence of the $\cp$-violating parameters $A_f$ and $M_3$.  
We assume a unit CKM matrix. 
Examples of genuine one-particle irreducible (1PI) vertex diagrams are shown in \reffi{diag}a,b,c.  
\reffi{diag}d shows an example of a reducible diagram, where a Higgs
boson mixes with a Z boson or a Goldstone boson.
For our calculations we have made use of the program \texttt{FeynArts},
allowing automated generation of the Feynman diagrams and
amplitudes~\cite{Kublbeck:1990xc,Hahn:2000kx,Hahn:2001rv}.  In
conjunction, we utilised the packages \texttt{FormCalc}  and \texttt{LoopTools} 
for the calculation of matrix elements and loop
integrals \cite{Hahn:1998yk}.  For regularisation we use dimensional
reduction, according to the prescription of
\citeres{Hahn:1998yk,delAguila:1998nd}. 
We supplemented the model files available in \texttt{FeynArts} with 
counterterms for the 2- and 3-point vertices involved, 
specified according to the renormalisation prescription outlined below.
\subsection{Renormalisation in the Chargino and Neutralino Sector}

A significant number of one-loop calculations have been carried out in the chargino--neutralino sector
 of the $\cp$-conserving MSSM
 with real parameters, see e.g.\ 
\citeres{Lahanas:1993ib,Pierce:1993gj,Pierce:1994ew,Eberl:2001eu,Fritzsche:2002bi,Oller:2003ge,Oller:2005xg, Drees:2006um}, 
with the renormalisation schemes of \citeres{Oller:2005xg,Rolbiecki:2007se} also applicable
 for complex parameters.  More recently, $\cp$-odd observables 
have been calculated at one-loop level in the $\cp$-violating MSSM, see
e.g.\ \citeres{Eberl:2005ay,Osland:2007xw,Rolbiecki:2007se},
 but these calculations did not always require a dedicated renormalisation scheme
  as the specific observables calculated were UV-finite.
  In order to renormalise the fields and parameters in this sector we introduce 
  counterterms and renormalisation constants of a similar 
form to \citere{Fritzsche:2002bi}.  However, we apply different
on-shell conditions for the mass parameters and we extend the formalism
to the general case including $\cp$-violation, properly taking into account
imaginary parts arising both from the complex MSSM parameters and from
absorptive parts of loop integrals.
In this work, we allow the sfermion trilinear couplings, $A_f$, and
 gluino mass parameter, $M_3$, to have $\cp$-violating phases, as 
inspired by the CPX scenario.  The full complex MSSM also contains phases
in the neutralino sector at tree level, where $M_1$, $M_2$ and $\mu$ may 
be complex (there are only two physical phases, since one of the phases 
of $M_1$ and $M_2$ can be rotated away). For the numerical results
obtained in
this paper we do not need to specify a renormalisation scheme
for these phases, since they are zero in the CPX scenario. A discussion of this issue will be deferred to a
forthcoming publication.
The mass matrices each receive a counterterm as follows,
\begin{eqnarray}
 X &\rightarrow X+ \delta X,\;\;\;\;\; Y &\rightarrow Y+ \delta Y ,
\end{eqnarray}
where $\delta X$ and $\delta Y$ are $2 \times 2$ and $4 \times 4$
matrices,
respectively.  Their elements contain three new renormalisation constants,
 $\delta M_1$, $\delta M_2$ and $\delta\mu$, as well as renormalisation constants from other sectors. 
We introduce renormalisation constants separately for the left and right-handed components of the incoming and outgoing fermion fields, as follows,
\begin{align}
  \omega_L \tilde{\chi}^-_i &\rightarrow (1+\tfrac{1}{2} \delta Z_-^L)_{ij} \omega_L  \tilde{\chi}^-_j,&
 \overline{\tilde{\chi}^{-}_i} \omega_R &\rightarrow \overline{\tilde{\chi}^-_i} (1+\tfrac{1}{2} \delta \bar{Z}_-^L)_{ij} \omega_R,  \nonumber\\
\omega_R \tilde{\chi}^-_i &\rightarrow (1+\tfrac{1}{2} \delta Z_-^{R})_{ij} \omega_R  \tilde{\chi}^-_j    
,& 
\overline{\tilde{\chi}^-_i} \omega_L &\rightarrow \overline{\tilde{\chi}^-_i} (1+\tfrac{1}{2} \delta \bar{Z}_-^R)_{ij}\omega_L,\nonumber\\
\omega_L \tilde{\chi}^0_i &\rightarrow (1+\tfrac{1}{2} \delta Z_0^L)_{ij} \omega_L  \tilde{\chi}^0_j,&
\overline{\tilde{\chi}^0_i} \omega_R &\rightarrow \overline{\tilde{\chi}^0_i} (1+\tfrac{1}{2} \delta \bar{Z}_0^L)_{ij} \omega_R,\nonumber\\
 \omega_R \tilde{\chi}^0_i &\rightarrow (1+\tfrac{1}{2} \delta Z_0^{R})_{ij} \omega_R  \tilde{\chi}^0_j  ,&
 \overline{\tilde{\chi}^0_i} \omega_L&\rightarrow
\overline{\tilde{\chi}^0_i} (1+\tfrac{1}{2} \delta \bar{Z}_0^R)_{ij}
\omega_L ,
\end{align}
where $j$ is summed over 1,2 (1,2,3,4) for the charginos (neutralinos). Note that we have introduced barred renormalisation constants 
for outgoing fermions and incoming antifermions.  In the $\cp$-conserving MSSM, these are related to the non-barred renormalisation
 constants for incoming fermions and outgoing antifermions by a Hermitian conjugate, i.e. $\delta \bar{Z}_{ij}=\delta Z^{\dagger}_{ij}$.  
For the $\cp$-violating MSSM, we choose to treat these quantities as independent at this stage, with more discussion to follow.  

 Inserting the above transformations into the Born Lagrangian,
 we obtain formulae for the renormalised 1PI two-point vertex functions, $\hat{\mathit{\Gamma}}_{ij}=i \hat{S}^{-1}_{ij}$,
where $\hat{S}_{ij}$ is the loop-corrected propagator and hatted quantities are renormalised.
The field renormalisation constants are then fixed by requiring
 that these $\hat{\mathit{\Gamma}}_{ij}$ are diagonal for on-shell external particle momenta and that the propagators have unity
 residues, namely,
 \begin{align}
\hat{\mathit{\Gamma}}_{ij} \tilde{\chi}_{i}(p)|_{p^2=m_{\tilde{\chi}_j}^2} &=0,& \overline{\tilde{\chi}}_{i}(p)\hat{\mathit{\Gamma}}_{ij}|_{p^2=m_{\tilde{\chi}_i}^2} &= 0,\label{eqn:nomixing}\\
\lim_{p^2 \rightarrow m_{\tilde{\chi}_i}^2} \frac{1}{\displaystyle{\not}p-m_{\tilde{\chi}_i}} \hat{\mathit{\Gamma}}_{ii} \tilde{\chi}_{i}(p) &= \tilde{\chi}_{i},&
\displaystyle{\lim_{p^2 \rightarrow m_{\tilde{\chi}_i}^2}}
\hat{\mathit{\Gamma}}_{ii} \overline{\tilde{\chi}}_{i}(p)
\frac{1}{\displaystyle{\not}p-m_{\tilde{\chi}_i}} &=
\overline{\tilde{\chi}}_{i} ,
\label{eqn:residues}
\end{align}
 where $\tilde{\chi}_{i}=\tilde{\chi}^-_{i}\;(i,j=1,2)$ or 
$\tilde{\chi}^0_{i}\;(i,j=1,2,3,4)$ and $i\neq j$.  
We also require that the loop-corrected propagator has the correct 
Lorentz structure in the on-shell limit. 
Namely, if we decompose the self-energies as 
\begin{equation}
 \Sigma_{ij}(p^2)=\displaystyle{\not}p\, \omega_L  \Sigma^L_{ij}(p^2)+\displaystyle{\not}p\, \omega_R  \Sigma^R_{ij}(p^2)
+\omega_L  \Sigma^{SL}_{ij}(p^2)+ \omega_R \Sigma^{SR}_{ij}(p^2)
\label{eqn:Lorentzse}
\end{equation}
then, in order to ensure $\hat{\mathit{S}}_{ii}$ has only a scalar and 
vector part on-shell, we require,
\begin{eqnarray}
 \hat{\Sigma}^{L}_{ii}(m_{\tilde{\chi}_i}^2)&=&
\hat{\Sigma}^{R}_{ii}(m_{\tilde{\chi}_i}^2) , \label{eqn:nochiralLR}\\
 \hat{\Sigma}^{SL}_{ii}(m_{\tilde{\chi}_i}^2)&=&
\hat{\Sigma}^{SR}_{ii}(m_{\tilde{\chi}_i}^2) \label{eqn:nochiralSLSR} .
\end{eqnarray}
The conditions in  \refeqs{eqn:nomixing}, (\ref{eqn:residues}) and 
(\ref{eqn:nochiralSLSR}) lead to the following off-diagonal and diagonal 
chargino field renormalisation constants, respectively 
(\refeq{eqn:nochiralLR} is then automatically satisfied).
\begin{eqnarray}
\delta Z^{L/R}_{-,ij}&=&\frac{2}{m_{\tilde{\chi}^{\pm}_i}^2-m_{\tilde{\chi}^{\pm}_j}^2}\big[m_{\tilde{\chi}^{\pm}_j}^2 \Sigma_{-,ij}^{L/R}(m_{\tilde{\chi}^{\pm}_j}^2)+m_{\tilde{\chi}^{\pm}_i}m_{\tilde{\chi}^{\pm}_j} \Sigma_{-,ij}^{R/L}(m_{\tilde{\chi}^{\pm}_j}^2)+m_{\tilde{\chi}^{\pm}_i}\Sigma_{-,ij}^{SL/SR}(m_{\tilde{\chi}^{\pm}_j}^2)\nonumber\\&&+m_{\tilde{\chi}^{\pm}_j}\Sigma_{-,ij}^{SR/SL}(m_{\tilde{\chi}^{\pm}_j}^2)
- m_{\tilde{\chi}^{\pm}_{i/j}}\big(U^\ast \delta X V^\dagger\big)_{ij}-
  m_{\tilde{\chi}^{\pm}_{j/i}}\big(V \delta X^\dagger U^T\big)_{ij}\big] , \\
\delta\bar{Z}^{L/R}_{-,ij}&=&\frac{2}{m_{\tilde{\chi}^{\pm}_j}^2-m_{\tilde{\chi}^{\pm}_i}^2} \big[m_{\tilde{\chi}^{\pm}_i}^2 \Sigma_{-,ij}^{L/R}(m_{\tilde{\chi}^{\pm}_i}^2)+m_{\tilde{\chi}^{\pm}_i}m_{\tilde{\chi}^{\pm}_j} \Sigma_{-,ij}^{R/L}(m_{\tilde{\chi}^{\pm}_i}^2)+m_{\tilde{\chi}^{\pm}_i}\Sigma_{-,ij}^{SL/SR}(m_{\tilde{\chi}^{\pm}_i}^2)\nonumber\\&&+m_{\tilde{\chi}^{\pm}_j}\Sigma_{-,ij}^{SR/SL}(m_{\tilde{\chi}^{\pm}_i}^2)
- m_{\tilde{\chi}^{\pm}_{i/j}}(U^\ast \delta X V^\dagger)_{ij}-
  m_{\tilde{\chi}^{\pm}_{j/i}}(V \delta X^\dagger U^T)_{ij}\big] ,\\
\delta Z^{L/R}_{-,ii}&=&-\Sigma_{-,ii}^{L/R}(m_{\tilde{\chi}^{\pm}_i}^2)\!-\!m_{\tilde{\chi}^{\pm}_i}^2
 \big[\Sigma_{-,ii}^{L^{\prime}}(m_{\tilde{\chi}^{\pm}_i}^2)\!+\!\Sigma_{-,ii}^{R^{\prime}}(m_{\tilde{\chi}^{\pm}_i}^2)\big]\!-\!
m_{\tilde{\chi}^{\pm}_i}
\big[\Sigma_{-,ii}^{SL^{\prime}}(m_{\tilde{\chi}^{\pm}_i}^2)\!+\!\Sigma_{-,ii}^{SR^{\prime}}(m_{\tilde{\chi}^{\pm}_i}^2)\big] \nonumber\\
 &&\pm\frac{1}{2
m_{\tilde{\chi}^{\pm}_i}}\big[\Sigma_{-,ii}^{SL}(m_{\tilde{\chi}^{\pm}_i}^2)\!-\!\Sigma_{-,ii}^{SR}(m_{\tilde{\chi}^{\pm}_i}^2)+(V
\delta X^{\dagger} U^{T})_{ii}\!-\!(U^{\ast}\delta X
V^{\dagger})_{ii}\big] , \\
\delta\bar{Z}^{L/R}_{-,ii}&=&-\Sigma_{-,ii}^{L/R}(m_{\tilde{\chi}^{\pm}_i}^2)\!-\!m_{\tilde{\chi}^{\pm}_i}^2
 \big[\Sigma_{-,ii}^{L^{\prime}}(m_{\tilde{\chi}^{\pm}_i}^2)\!+\!\Sigma_{-,ii}^{R^{\prime}}(m_{\tilde{\chi}^{\pm}_i}^2)\big]\!-\! m_{\tilde{\chi}^{\pm}_i} \big[\Sigma_{-,ii}^{SL^{\prime}}(m_{\tilde{\chi}^{\pm}_i}^2)\!+\!\Sigma_{-,ii}^{SR^{\prime}}(m_{\tilde{\chi}^{\pm}_i}^2)\big]\nonumber\\
 &&\mp\frac{1}{2
m_{\tilde{\chi}^{\pm}_i}}\big[\Sigma_{-,ii}^{SL}(m_{\tilde{\chi}^{\pm}_i}^2)\!-\!\Sigma_{-,ii}^{SR}(m_{\tilde{\chi}^{\pm}_i}^2)+(V
\delta X^{\dagger} U^{T})_{ii}\!-\!(U^{\ast}\delta X
V^{\dagger})_{ii}\big] , 
\end{eqnarray}
while the field renormalisation constants for neutralinos are given by
\begin{eqnarray}
\delta Z^{L/R}_{0,ij}=\delta\bar{Z}^{R/L}_{0,ji}\!\!\!\!&=&\!\!\!\!\frac{2}{m_{\tilde{\chi}^0_i}^2-m_{\tilde{\chi}^0_j}^2}\big[m_{\tilde{\chi}^0_j}^2 \Sigma_{0,ij}^{L/R}(m_{\tilde{\chi}^0_j}^2)+m_{\tilde{\chi}^0_i}m_{\tilde{\chi}^0_j} \Sigma_{0,ij}^{R/L}(m_{\tilde{\chi}^0_j}^2)+m_{\tilde{\chi}^0_i}\Sigma_{0,ij}^{SL/SR}(m_{\tilde{\chi}^0_j}^2)\nonumber\\&&+m_{\tilde{\chi}^0_j}\Sigma_{0,ij}^{SR/SL}(m_{\tilde{\chi}^0_j}^2)
- m_{\tilde{\chi}^0_{i/j}}\big(N^\ast \delta Y N^\dagger\big)_{ij}-
  m_{\tilde{\chi}^0_{j/i}}\big(N \delta Y^\dagger N^T\big)_{ij}\big] ,\\
\delta Z^{L/R}_{0,ii}=\delta\bar{Z}^{R/L}_{0,ii}\!\!\!\!&=&\!\!\!\!-\Sigma_{0,ii}^{L/R}(m_{\tilde{\chi}^0_i}^2)\!-\!m_{\tilde{\chi}^0_i}^2
 \big[\Sigma_{0,ii}^{L^{\prime}}(m_{\tilde{\chi}^0_i}^2)\!+\!\Sigma_{0,ii}^{R^{\prime}}(m_{\tilde{\chi}^0_i}^2)\big]\!-\! m_{\tilde{\chi}^0_i} \big[\Sigma_{0,ii}^{SL^{\prime}}(m_{\tilde{\chi}^0_i}^2)\!+\!\Sigma_{0,ii}^{SR^{\prime}}(m_{\tilde{\chi}^0_i}^2)\big]\nonumber\\
 &&\pm\frac{1}{2
m_{\tilde{\chi}^0_i}}\big[\Sigma_{0,ii}^{SL}(m_{\tilde{\chi}^0_i}^2)\!-\!\Sigma_{0,ii}^{SR}(m_{\tilde{\chi}^0_i}^2)+(N
\delta Y^{\dagger} N^{T})_{ii}\!-\!(N^{\ast}\delta Y
N^{\dagger})_{ii}\big] .
\end{eqnarray}
Here $\Sigma_{ii}^{\prime}(m_i^2)$ denotes the derivative 
$\frac{\partial \Sigma_{ii}(p^2)}{\partial p^2}|_{p^2=m_i^2}$, 
and $\Sigma_-$ and $\Sigma_0$ indicate chargino and neutralino
self-energies, respectively.  The
relations between the left- and right-handed constants for the Majorana neutralinos result from charge conjugation symmetry.   
For both the charginos and neutralinos, the barred constants, 
$\delta\bar{Z}^{L/R}_{ij}$, differ from 
 $(\delta Z^{L/R}_{ij})^{\dagger}$ in their absorptive parts only.  
In the $\cp$-conserving MSSM, this difference vanishes.  
Also, the $\frac{1}{2 m_{\tilde{\chi}_i}}$ 
terms in the diagonal constants vanish if there are no complex
parameters, and we recover the formulae from
Ref.~\cite{Fritzsche:2002bi} in this case.
In the $\cp$-violating MSSM, this term is non-zero (this term also 
appears as a purely imaginary contribution in Ref.~\cite{Fritzsche:thesis}, 
with which our results agree up to absorptive parts).  
In the $\cp$-violating MSSM, the absorptive parts 
of loop integrals for unstable particles may
 enter the squared matrix element at the one-loop level since they can be 
multiplied by imaginary coefficients arising
 from the complex parameters.
In the literature, the issue of the treatment of absorptive parts of
loop integrals in field renormalisation constants 
has found considerable attention, mostly in the context
of the renormalisation of the SM, see e.g.\
\citeres{Denner:1991kt,Espriu:2000fq,Espriu:2002xv,
Denner:2004bm,Zhou:2005mc,Kniehl:2009kk}. A possibility that has been
advocated for instance in \citeres{Denner:1991kt,Denner:2004bm} is to
discard the absorptive parts of loop integrals in the field 
renormalisation constants,
while keeping any complex parameters in the coefficients, 
indicated by inserting the symbol $\ReTilde$ 
into the renormalisation conditions in 
 \refeqs{eqn:nomixing}, (\ref{eqn:residues}), (\ref{eqn:nochiralLR}) and 
(\ref{eqn:nochiralSLSR}).
With this choice the hermiticity relation, 
$\delta\bar{Z}^{L/R}_{ij}= (\delta Z^{L/R}_{ij})^{\dagger}$,
is restored,
but one must include all reducible self-energy diagrams and may 
have to introduce additional finite
 normalisation constants to ensure the external particles have the correct 
on-shell properties. Renormalisation conditions without 
$\ReTilde$ were used in Ref.~\cite{Espriu:2002xv} for the SM, as a way 
of ensuring the correct on-shell conditions and
gauge-independent matrix elements. Although the hermiticity relation 
between renormalisation constants is not valid in this case, the authors of
Ref.~\cite{Espriu:2002xv} showed that the CPT theorem still holds.  
Nevertheless, the issue of an appropriate field renormalisation 
of unstable particles on external legs remains under debate in the literature.  
For the class of processes considered in this paper, it turns out that all 
absorptive parts of external neutralino
self-energy diagrams cancel when the squared matrix element is summed over 
all spins.  This is due to the relation between left-
and right-handed components of the (Majorana) neutralinos, 
and does not apply for (Dirac) charginos nor for spin-dependent 
calculations.  
Hence, for the numerical results presented in this paper absorptive
parts of loop integrals do not contribute.

It should be noted that the prescription for the field renormalisation
constants given above is valid for the most general case of
$\cp$-violating parameters in the complex MSSM. As mentioned above, 
we restrict the analyses in this paper to cases where the 
parameters $M_1$, $M_2$ and $\mu$ are real. Therefore we do not specify
the renormalisation of $\cp$-violating phases of those parameters in what follows below. This issue
will be addressed in a forthcoming publication.

For the parameter renormalisation of $M_1$, $M_2$, $\mu$, we use an 
on-shell approach, because this is convenient in processes with 
external charginos and neutralinos.  
In the chargino--neutralino sector, we have three independent input 
parameters, $M_1$, $M_2$, $\mu$,
which determine the tree-level masses, $m_{\widetilde{\chi}_i}$, of the 
six fields, $\tilde{\chi}^\pm_{1,2}$, $\tilde{\chi}^0_{1,2,3,4}$.
  The loop-corrected masses, $M_{\widetilde{\chi}_i}$, are then defined 
as the real parts of the poles
of the corresponding loop-corrected propagators $\hat{S}_{ii}$.  
At one-loop order they may be written in terms of the renormalised 
self-energies as follows,
\begin{equation}
 M_{\widetilde{\chi}_i}=m_{\widetilde{\chi}_i} [1-\frac{1}{2}\mathrm{Re}[\hat{\Sigma}^L_{ii}(m_{\widetilde{\chi}_i}^2)+\hat{\Sigma}^{R}_{ii}(m_{\widetilde{\chi}_i}^2)]]-\frac{1}{2}\mathrm{Re}[\hat{\Sigma}^{SL}_{ii}(m_{\widetilde{\chi}_i}^2)+\hat{\Sigma}^{SR}_{ii}(m_{\widetilde{\chi}_i}^2)].
\label{masses1}
\end{equation}
We fix three of these masses on-shell by requiring that the pole masses $M_{\widetilde{\chi}_i}$ coincide 
with their tree level values $m_{\widetilde{\chi}_i}$.  This gives us three equations to solve for $\delta M_1$, $\delta M_2$ and $\delta \mu$.
  The remaining three masses will be different to their tree-level values.  
There is obviously a freedom of choice here in the three masses that are
used in the on-shell conditions. It should be noted that the ``most
convenient'' choice for those masses will depend on the process under
consideration and may even be different in different regions of
parameter space. In Ref.~\cite{Fritzsche:2002bi}, 
the masses of $\tilde{\chi}^0_{1},\,\tilde{\chi}^\pm_{1}$ and 
$\tilde{\chi}^\pm_{2}$ were fixed on-shell.
  This choice is advantageous for processes where charginos appear 
as external particles, and ensures a proper cancellation of the 
infra-red divergences present in QED corrections.
For the processes considered in the present paper, 
it is convenient to have the two lightest neutralinos on-shell.
  We thus choose to fix the masses of  
$\tilde{\chi}^0_{1}$, $\tilde{\chi}^0_{2}$ 
and $\tilde{\chi}^\pm_{2}$ on-shell.  We found this to give numerically stable 
results for the hierarchy of $M_1<M_2\ll\mu$ among the mass parameters
in the chargino--neutralino sector, while for other processes and 
parameters we found that different choices can be favourable.  The resulting 
expressions for $\delta M_1$, $\delta M_2$ and $\delta \mu$ are given
below,
\begin{eqnarray}
  \delta M_1&=&[2 (\ZNeu_{13} \ZNeu_{14}\ZNeu_{22}^2 -\ZNeu_{12}^2\ZNeu_{23}
      \ZNeu_{24})C_{(2)}
    + (\UCha_{22}\VCha_{22} \ZNeu_{22}^2+2 \UCha_{21}\VCha_{21}\ZNeu_{23}\ZNeu_{24})N_{(1)}\nonumber\\&&
     -(\UCha_{22}\VCha_{22}\ZNeu_{12}^2+2
\UCha_{21}\VCha_{21}\ZNeu_{13}\ZNeu_{14})N_{(2)}]/K , \\
 \delta M_2&=&[2(\ZNeu_{11}^2\ZNeu_{23}\ZNeu_{24}-\ZNeu_{13}\ZNeu_{14}\ZNeu_{21}^2)C_{(2)} - \UCha_{22}\VCha_{22}\ZNeu_{21}^2 N_{(1)} -
 \UCha_{22}\VCha_{22}\ZNeu_{11}^2N_{(2)}]/K , \\
 \delta \mu&=&[-(\ZNeu_{12}^2\ZNeu_{21}^2- \ZNeu_{11}^2\ZNeu_{22}^2) C_{(2)}+ 
   \UCha_{21}\VCha_{21}\ZNeu_{21}^2
N_{(1)}-\UCha_{21}\VCha_{21}\ZNeu_{11}^2 N_{(2)}]/K ,
\end{eqnarray}
where
\begin{eqnarray}
C_{(i)}&=&\mathrm{Re}
\big[m_{\tilde{\chi}^\pm_i}[\Sigma^L_{ii}(m_{\tilde{\chi}^\pm_i}^2)+
\Sigma^R_{ii}(m_{\tilde{\chi}^\pm_i}^2)]+
      \Sigma^{SL}_{ii}(m_{\tilde{\chi}^\pm_i}^2)+\Sigma^{SR}_{ii}(m_{\tilde{\chi}^\pm_i}^2)\big]\nonumber\\&&- 
     2\delta X_{21} \UCha_{i2}\VCha_{i1}- 2\delta
X_{12}\UCha_{i1}\VCha_{i2} , \nonumber \\
N_{(i)}&=&\mathrm{Re} \big[m_{\tilde{\chi}^0_i}[\Sigma^L_{ii}(m_{\tilde{\chi}^0_i}^2)+\Sigma^R_{ii}(m_{\tilde{\chi}^0_i}^2)] +\Sigma^{SL}_{ii}(m_{\tilde{\chi}^0_i}^2)+ \Sigma^{SR}_{ii}(m_{\tilde{\chi}^0_i}^2)\big] \nonumber\\&&- 4\delta Y_{13}\ZNeu_{i1}\ZNeu_{i3} - 
       4\delta Y_{23}\ZNeu_{i2}\ZNeu_{i3} - 4\delta Y_{14}\ZNeu_{i1}
        \ZNeu_{i4} - 4\delta Y_{24}\ZNeu_{i2}\ZNeu_{i4} , \nonumber\\
 K&=& 2\UCha_{22}\VCha_{22}(\ZNeu_{11}^2\ZNeu_{22}^2-\ZNeu_{12}^2\ZNeu_{21}^2) + 4\UCha_{21}\VCha_{21}
    (\ZNeu_{11}^2\ZNeu_{23}\ZNeu_{24}-\ZNeu_{13}\ZNeu_{14}\ZNeu_{21}^2) .
\end{eqnarray}
Note that, since we are assuming $M_1$, $M_2$ and $\mu$ to be real, 
the mixing matrix elements always appear in combinations where the conjugate is not needed.
\subsection{Renormalisation in the Higgs Sector}
For the Higgs sector we follow the renormalisation scheme of 
Ref.~\cite{Frank:2006yh}.  The independent
 parameters of the Higgs sector are taken to be $M_{H^{\pm}}$ and $\tan{\beta}$.  One 
field renormalisation constant is introduced for each Higgs doublet,
and $\tan{\beta}$ receives a counterterm as follows
\begin{equation} \mathcal{H}_{1,2} \rightarrow (1+\frac{1}{2}\delta Z_{\mathcal{H}_{1,2}})\mathcal{H}_{1,2}, 
\;\;\;\;\;\;\tan{\beta} \rightarrow \tan{\beta}(1+\delta \tan{\beta}).\end{equation} 
As in \citere{Frank:2006yh} we adopt $\overline{\mathrm{DR}}$ renormalisation for the fields and $\tan{\beta}$,
where the counterterm for the latter is given by
$\delta \tan{\beta}^{\overline{\mathrm{DR}}}=
\frac{1}{2}(\delta Z^{\overline{\mathrm{DR}}}_{\mathcal{H}_2}
-\delta Z^{\overline{\mathrm{DR}}}_{\mathcal{H}_1})$,
and on-shell renormalisation of the charged Higgs boson mass,
\begin{equation}
 \delta M_{H^{\pm}}^2 = \mathrm{Re}\;\Sigma_{H^+H^-}(M_{H^{\pm}}^2).
\end{equation}
The loop-corrected neutral masses $M_{h_a}$ are then defined as the real parts of the poles of the diagonal
 elements of the $3 \times 3$ Higgs propagator matrix, as in Ref.~\cite{Frank:2006yh}, with $M_{h_1}\leq M_{h_2}\leq M_{h_3}$.

The correct on-shell properties of Higgs bosons appearing as external 
particles in physical processes, and thus a properly 
normalised S-matrix, are ensured by the introduction of finite
wavefunction normalisation factors $\hat{Z}_{ij}$. These Z-factors
are a convenient way to account for the
 mixing between the Higgs bosons and to incorporate leading higher-order
contributions.  Following Ref.~\cite{Frank:2006yh}, a renormalised 1PI vertex
 $\hat{\mathit{\Gamma}}_i$ with an external Higgs boson $i$ ($i=h,H,A)$
then takes the form
\begin{equation}
\sqrt{\hat{Z_i}}(\hat{\mathit{\Gamma}}_i+\hat{Z}_{ij}\hat{\mathit{\Gamma}}_j+\hat{Z}_{ik}\hat{\mathit{\Gamma}}_k+...).
\label{zfactors}
\end{equation}
Here $j,k$ are the remaining two of $h,H,A$ and are not summed over, and the ellipsis refers to the mixing contributions 
with the Goldstone and Z bosons which we will consider in Section \ref{sec_vertexrenorm}.  Without the ellipsis, the normalisation of 
the wavefunctions can be expressed in terms of a $3 \times 3$ non-unitary matrix $\mathbf{\hat{Z}}$, where
 $\mathbf{\hat{Z}}_{ij}\equiv\sqrt{\hat{Z_i}}\hat{Z}_{ij}$ and $\hat{Z}_{ii}=1$.  We use the same formulae 
for the $\mathbf{\hat{Z}}$ matrix elements in terms of the Higgs self-energies as those derived in Ref. \cite{Williams:2007dc}.

\subsection{Renormalisation in Other Sectors}

We parameterise the electric charge, $e=\sqrt{4 \pi \alpha}$, in terms of $\alpha (M_Z)=\alpha(0)/(1-\Delta\alpha)$,
 where 
$\Delta\alpha=\Delta \alpha_{\rm lept}+ \Delta \alpha_{\rm had}^{(5)}$ is the shift in the fine-structure constant arising
 from large logarithms of light fermions.  This yields the following
counterterm for the charge renormalisation,
\begin{equation}
\label{eq:chargeren}
\delta Z_e=\frac{1}{2}
\Pi_{\gamma}(0)-\frac{s_W}{c_W}\frac{\Sigma_{\gamma
Z}^T(0)}{M_Z^2}-\frac{\Delta\alpha}{2} .
\end{equation}
Here $\Pi_{\gamma}(0)=\frac{\partial
\Sigma_{\gamma\gamma}(k^2)}{\partial k^2}\rvert_{k^2=0}$ is the photon
vacuum polarisation, and the large logarithms  involving light fermion
masses drop out in \refeq{eq:chargeren}.
The renormalisation constants in the gauge boson sector are defined as follows,
\begin{eqnarray}
M_Z^2 \rightarrow M_Z^2+\delta M_Z^2, \;\;\;\; M_W^2 \rightarrow
M_W^2+\delta M_W^2, \;\;\;\;s_W \rightarrow s_W+\delta s_W ,
\end{eqnarray}
where 
\begin{equation}
\delta s_W = \frac{c_W^2}{2 s_W} \left(\frac{\delta
M_Z^2}{M_Z^2}-\frac{\delta M_W^2}{M_W^2}\right) .
\end{equation}
The renormalisation constants are then deduced from on-shell conditions for the masses of the W and Z bosons,
\begin{equation}
 \delta M_{W}^2 = \mathrm{Re}\;\Sigma^T_{WW}(M_{W}^2), \;\;\;\; \delta
M_{Z}^2 = \mathrm{Re}\;\Sigma^T_{ZZ}(M_{Z}^2).        
\end{equation}  

\subsection{Vertex Renormalisation}
\label{sec_vertexrenorm}

The 3-point vertex for $\tilde{\chi}^0_i\tilde{\chi}^0_j h^0_k$, where $h^0_k=\{h,H,A,G\}$, 
can be renormalised by a coupling counterterm as follows,
\begin{eqnarray}
\delta C^{R/L}_{ijh_k^0} &=&\frac{e}{2 c_W s_W} \delta c^{(\ast)}_{ijh_k^0}+ C^{R/L}_{ijh_k^0}(\delta Z_e-\frac{\delta s_W}{s_W}
-\frac{\delta c_W}{c_W})+\frac{1}{2}\sum^4_{l=1}(\delta Z^{R/L}_{li} C^{R/L}_{ljh_k^0}+
\delta \bar{Z}^{L/R}_{jl} C^{R/L}_{ilh_k^0})\nonumber\\
&&+\frac{1}{2}(\delta Z_{h_k^0 h} C^{R/L}_{ijh}+ \delta Z_{h_k^0 H}C^{R/L}_{ijH}+\delta Z_{h_k^0 A}C^{R/L}_{ijA}+\delta Z_{h_k^0 G}C^{R/L}_{ijG})   
\end{eqnarray}
where
\begin{equation}
\delta c_{ijh^0_k} =[(a_k N_{i3} + b_k N_{i4})(\delta s_W N_{j1} - \delta c_W N_{j2})+(a_k N_{j3}+b_k N_{j4})
(\delta s_W N_{i1} - \delta c_W N_{i2})].
\end{equation}
The 3-point vertex for $\tilde{\chi}^0_i\tilde{\chi}^0_j h_a$ is then
constructed using the $3 \times 3$ $\mathbf{\hat{Z}}$ matrix
 for the normalisation of wavefunctions as in \refeq{zfactors}.  This 
automatically includes the reducible self-energy 
diagrams involving $h,H,A$. For a complete one-loop
result, reducible diagrams
 involving mixing self-energies of Higgs bosons with the G and Z bosons, 
such as those in Figure \ref{diag}(d), must also be included.  In order to  
ensure a proper cancellation of the gauge parameter dependence,
 we follow the approach of Ref.~\cite{Williams:2007dc} and evaluate these reducible contributions, $\hat{\mathit{\Gamma}}^{\mathrm{G,Z.se}}$,
 strictly at the one-loop level. Our full result,
$\hat{\mathit{\Gamma}}^{\mathrm{Full\,Loop}}$ is then obtained by
combining these contributions with those of genuine vertex type, 
$\hat{\mathit{\Gamma}}^{\mathrm{1PI}}$, as follows,
\begin{equation}
\hat{\mathit{\Gamma}}^{\mathrm{Full\,Loop}}_{\tilde{\chi}^0_{i}\tilde{\chi}^0_{j}
 h_a}=\mathbf{\hat{Z}}_{al}[\hat{\mathit{\Gamma}}^{\mathrm{1PI}}_{\tilde{\chi}^0_{i}\tilde{\chi}^0_{j}
 h_l^0}(M_{h_a}^2)+\hat{\mathit{\Gamma}}^{\mathrm{G,Z.se}}_{\tilde{\chi}^0_{i}\tilde{\chi}^0_{j}
h_l^0}(m_{h^0_l}^2)] ,
\label{eqamp}
\end{equation}
where $h^0_l=\{h,H,A\}$ are the tree-level states with tree-level masses, $m_{h^0_l}$, and are summed over. 
 In contrast, $M_{h_a}$ is the loop-corrected mass of the Higgs boson $h_a$ in the physical
 process, i.e.\ one of $h_1, h_2, h_3$.  
Numerically, inclusion of the G--Z mixing did not have a significant effect. 
Across the CPX parameter space studied, the effect of this correction on the decay widths 
was less than $0.1\%$.

\subsection{Combination with Higher-order Results}
\label{sec_higherorder}

As Higgs propagator-type corrections are known to be large, we have
combined our one-loop result for the genuine vertex contribution 
with state-of-the-art two-loop propagator-type corrections obtained
within the Feynman diagrammatic approach, as implemented in the program
\texttt{FeynHiggs}~\cite{feynhiggs,fhrandproc,Frank:2006yh,mhcpv2l}.
These contributions incorporate in particular the full phase dependence
at $\mathcal{O}(\alpha_t \alpha_s)$, while we do not include here
further two-loop corrections that are known only for the case of real
MSSM parameters. Using \refeq{eqamp}, 
we combine the two-loop $\mathbf{\hat{Z}}$ factors and Higgs masses 
$M_{h_a}$ from \texttt{FeynHiggs 2.6.5}, with our own genuine
 vertex ($\hat{\mathit{\Gamma}}^{\mathrm{1PI}}$) and G--Z mixing ($\hat{\mathit{\Gamma}}^{\mathrm{G,Z.se}}$) corrections to the process
 $\tilde{\chi}^0_i\rightarrow \tilde{\chi}^0_j h_a$, thereby obtaining the most precise predictions for the corresponding 
decay widths and branching ratios in the MSSM with complex parameters.

In order to investigate the effects of the genuine vertex contributions
for the process $\tilde{\chi}^0_i\rightarrow \tilde{\chi}^0_j h_a$
we will in the following compare our full result with an Improved Born
approximation. The latter is obtained by summing over the tree-level 
amplitudes for 
$\tilde{\chi}^0_i\rightarrow \tilde{\chi}^0_j h^0_k$, weighted by the appropriate $\mathbf{\hat{Z}}$ factors and 
evaluated at the loop-corrected Higgs masses,
\begin{equation}
\hat{\mathit{\Gamma}}^{\mathrm{Improved\,Born}}_{\tilde{\chi}^0_{i}\tilde{\chi}^0_{j}
 h_a}=\mathbf{\hat{Z}}_{al}[\hat{\mathit{\Gamma}}^{\mathrm{Born}}_{\tilde{\chi}^0_{i}\tilde{\chi}^0_{j} h_l}(M_{h_a^2})].
\label{eqnib}
\end{equation}
We will always compare our numerical results to this Improved Born approximation, rather than to 
the strict tree-level result of \refeq{eqntree}.  This allows us to separate out the effect of our new genuine
 (process-specific) vertex corrections from those corrections coming from mixing effects and mass shifts in the Higgs 
sector which are already known to be large.  Thus when we speak of the percentage effect of our genuine vertex loop 
calculations on the partial decay width, $\Gamma$, we are referring to the ratio
\begin{equation}
\label{eq:rratio}
r=\frac{\Gamma_{\mathrm{Full\;Loop}}-\Gamma_{\mathrm{Improved\,Born}}}{\Gamma_{\mathrm{Improved\,Born}}}.
\end{equation}
As well as our full MSSM calculation, we will show approximations, where
only some (UV-finite)
 sets of diagrams such as third generation quarks and squarks, i.e.\
$t,\tilde{t},b,\tilde{b}$, are 
 included in the genuine vertex corrections.  In all cases, the
 two-loop propagator-type corrections from \texttt{FeynHiggs} are
evaluated in the full MSSM.  Various other approximations
 exist in the literature.  In Ref.~\cite{Zhang:2002fu}, only the one-loop 3rd generation (s)quark contributions in the real MSSM were
 considered.  In Ref.~\cite{Eberl:2001vb}, all one-loop (s)fermion contributions in 
the real MSSM were considered.  
Our full results thus go beyond these works, as we 
include all possible MSSM particles in the loops, we allow complex trilinear coupling and gluino parameters, and
 we incorporate complete one-loop and leading two-loop contributions
from the Higgs sector.

\section{Numerical Results for the Decay Width}
\label{sectionwidth}

As explained above, we will discuss in particular the
case of the CPX benchmark scenario~\cite{Carena:2000ks}, which gives rise to 
an unexcluded parameter region with a light Higgs. We use the following 
parameters for the CPX scenario unless specified otherwise,
\begin{eqnarray}
\label{cpxnumbers}
\mbox{CPX: } &&
\mu=2\,\mathrm{TeV}, \; M_{\mathrm{SUSY}}=500\,\mathrm{GeV}, \;
\vert M_3\vert=1\,\mathrm{TeV},\; 
\vert A_{f}\vert=900\,\mathrm{GeV},\nonumber\\
&& \phi_{M_3}\!=\!\phi_{A_{\mathrm{t,b},\tau}}\!=\!\pi/2, \;
M_2=200\,\mathrm{GeV},\; M_1=(5/3) t_W^2 M_2,\;
m_t=172.4\,\mathrm{GeV},
\end{eqnarray}
where $M_{\mathrm{SUSY}} = M_{L} = M_{\tilde{f}_R}$, see
\refeq{sfermion}, and $t_W \equiv s_W/c_W$. The values given in
\refeq{cpxnumbers} differ from the ones given in \citere{Carena:2000ks} 
in the value of the top-quark mass and in value of $\vert A_{f}\vert$,
for which an on-shell value is used that is slightly shifted from the 
$\overline{\mathrm{DR}}$ value specified in \citere{Carena:2000ks} (see also
\citere{Williams:2007dc}). In \refeq{cpxnumbers} 
we specify $M_{\mathrm{SUSY}}$ and $\vert A_{f}\vert$ for all sfermions, 
although it is only the third generation 
which plays a significant role in Higgs phenomenology.  For sfermions in the first and second generations, EDM constraints are more
 stringent (see \citere{Ellis:2008zy} for a recent study and
references therein), so we set the corresponding phases of $A_f$ to zero
in the first two generations.
The value of $M_2$ does not play a large role in Higgs
 phenomenology; we choose the nominal value of $M_2=200\,\mathrm{GeV}$
in order to agree with other studies, but we will investigate how
 its variation affects our results.   We will also show the effect of varying $\mu$ and $A_{\mathrm{t,b},\tau}$.  The masses 
of the supersymmetric particles in the CPX scenario of \refeq{cpxnumbers} are given in Table \ref{tab:masses}.
\begin{table}
\begin{tabular}{|c|c|c|c|c|c|c|c|}
\hline
$M_{\widetilde{\chi}^0_{3,4},\widetilde{\chi}^+_2}$ & $M_{\widetilde{g}}$ & $M_{\widetilde{u},\widetilde{d},\widetilde{c},\widetilde{s}}$ & $M_{\widetilde{t}_{1,2}}$ & $M_{\widetilde{b}_{1,2}}$ &  $M_{\widetilde{\chi}^0_2,\widetilde{\chi}^+_1}$ & $M_{\widetilde{\chi}^0_1}$ \\\hline
$\simeq$2001.2(-5),2003.1(-2),2003.4(1) & 1000 & $\simeq$500 & 332,667 & 471,531 & 198.5,198.5  & 94.7\\\hline
\end{tabular}
\caption{\label{tab:masses} \footnotesize{Masses in GeV of sparticles in
the CPX scenario with $\tan{\beta}=5.5$, where the neutralino
 and chargino masses are the tree-level values used in the loops.  The
number in brackets is the loop-correction to the last digit, evaluated
from \refeq{masses1}.}}
\end{table}
For comparison, we also present numerical results for the 
$\cp$-conserving case. In particular, we consider 
the ``small $\alpha_{\mathrm{eff}}$ scenario''~\cite{smallaleff}, with
the following parameter values
\begin{eqnarray}
\mbox{small } \alpha_{\mathrm{eff}}: &&
\mu=2\,\mathrm{TeV},\; M_{\mathrm{SUSY}}=800\,\mathrm{GeV},\;
\vert M_3\vert=500\,\mathrm{GeV},\;X_{f}=-1.1\,\mathrm{TeV}\nonumber\\
&& M_2=500\,\mathrm{GeV},\;M_1=(5/3) t_W^2 M_2,\;m_t=172.4\,\mathrm{GeV}.
\end{eqnarray}
The small $\alpha_{\mathrm{eff}}$ scenario has some similarities to the
CPX scenario, including large values of $\mu$ and $\vert A_f \vert$,
where $A_f$ is related to $X_f$ according to \refeq{Xf}.  As usual for these
benchmark scenarios, the parameters characterising the Higgs sector
at lowest order, i.e.\ $\MHp$ ($\MA$) and $\tb$ in the case of the CPX 
(small $\alpha_{\mathrm{eff}}$) scenario, are varied.

We furthermore investigate a specific case of a $\cp$-conserving scenario 
giving rise to a
very light $\tilde{\chi}^0_1$,
inspired by a recent study \cite{Dreiner:2009ic} which showed that very
light neutralinos are not ruled out by experimental data.  Here the GUT
relation between $M_1$ and $M_2$ is relaxed, allowing $M_1$ to be chosen
to be such that the lightest neutralino is approximately massless.  
Unless stated otherwise we use the following parameters,
\begin{eqnarray}
\mbox{light } \tilde{\chi}^0_1: &&
\mu=600\,\mathrm{GeV},\; M_{\mathrm{SUSY}}=500\,\mathrm{GeV},\;
\vert M_3\vert=1\,\mathrm{TeV},\;\tan\beta=20\nonumber\\
&& A_{f}=1\,\mathrm{TeV},\;M_2=400\,\mathrm{GeV},\;m_{\tilde{\chi}^0_1}=0,\;
m_t=172.4\,\mathrm{GeV}.
\label{eqnlightchi1}
\end{eqnarray}

\begin{figure}[htb!]
\centering
\subfigure[]{\label{Mh111}\includegraphics[height=7.2cm]{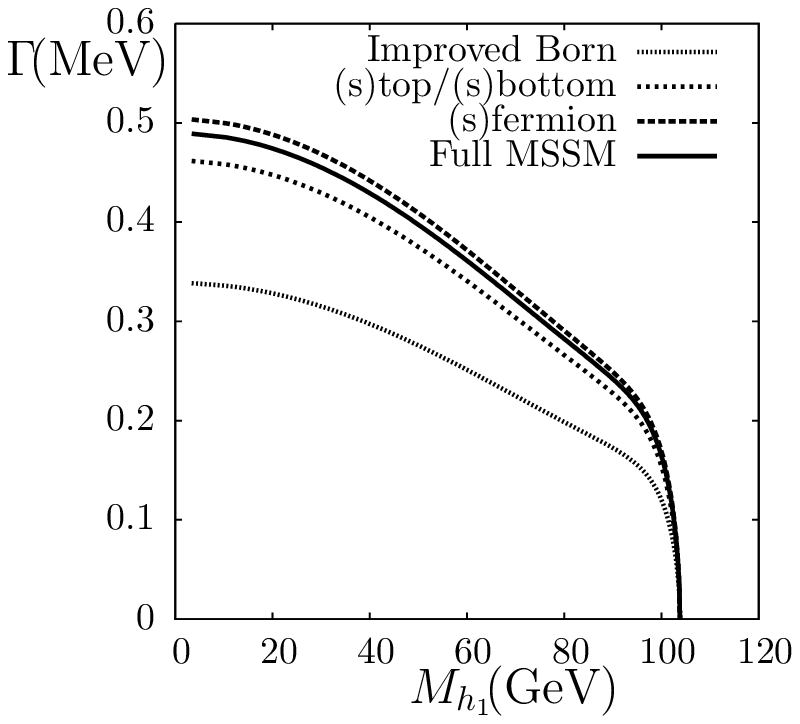}}
\subfigure[]{\label{Mh1ratio}\includegraphics[height=7.2cm]{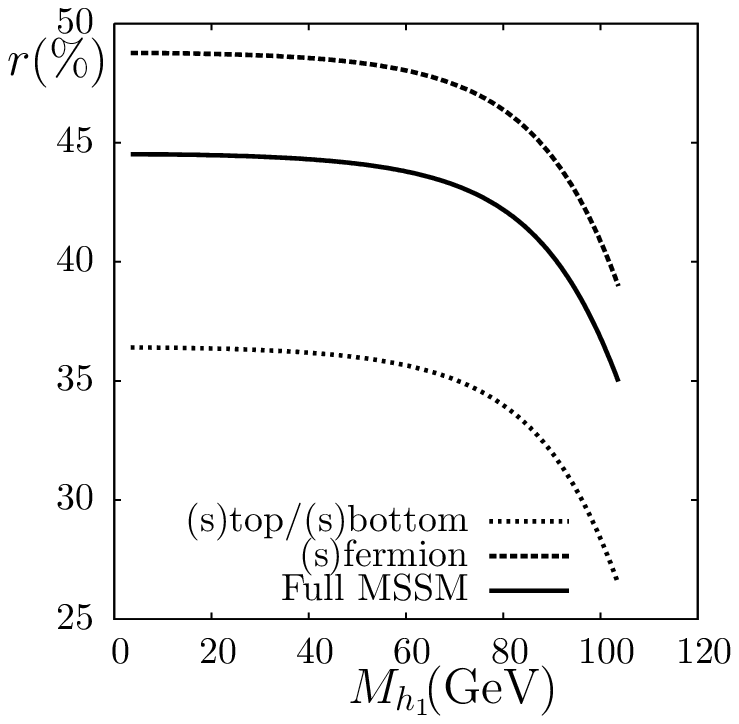}}
\caption{\footnotesize{Results for (a) the decay width 
$\Gamma(\tilde{\chi}^0_2\rightarrow\tilde{\chi}^0_1 h_1)$ and 
(b) the ratio
$r=(\Gamma_{\mathrm{Full\;Loop}}-\Gamma_{\mathrm{Improved\,Born}})/
\Gamma_{\mathrm{Improved\,Born}}$ 
in the 
CPX scenario plotted against $M_{h_1}$ for $\tan{\beta}=5.5$. ($M_{H^{\pm}}$ was varied as input.) 
The different curves indicate the inclusion of various subsets of diagrams.}}
\end{figure}

We start with numerical results for our genuine vertex corrections 
to the decay width for $\tilde{\chi}^0_2\rightarrow\tilde{\chi}^0_1 h_1$
 in the CPX scenario.  Figure \ref{Mh111} shows the partial decay width 
$\Gamma(\tilde{\chi}^0_2\rightarrow\tilde{\chi}^0_1 h_1)$ as a function of $M_{h_1}$.
  The value of $\tan{\beta}$ is fixed at $5.5$, while $M_{H^{\pm}}$ is varied as input.  
The dotted Improved Born curve shows the result obtained by combining
the tree-level amplitudes with 2-loop $\mathbf{\hat{Z}}$ matrix elements
and masses according to \refeq{eqnib}.
The other curves incorporate our new results for the genuine vertex
corrections, taking into account different sets of loop contributions.
Figure \ref{Mh1ratio} shows the ratio $r$, defined in \refeq{eq:rratio}, 
of the genuine vertex corrections relative to the Improved Born result as a function of $M_{h_1}$.  
We see from the figure that the impact of the genuine vertex corrections
on the decay width is very large. The corrections from the full
MSSM contributions to the vertex amount to about 45\% for Higgs
mass values in the region of the ``CPX hole'', i.e.\ for
$M_{h_1} \sim 40\,\mathrm{GeV}$. As expected,
the dominant effect arises from 
the triangle diagrams containing third generation quarks and squarks 
($t,\tilde{t},b,\tilde{b}$),
 due to the large top Yukawa coupling, yielding a correction of about
35\% compared to the Improved Born result. The other (s)fermions also
play a non-negligible role, in particular through their couplings to
neutralinos, increasing the total (s)fermion
 contribution to just under $50\%$.  The vertex corrections from the 
remainder of the particles in the MSSM, namely the vector bosons, Higgs
bosons, neutralinos and charginos, are negative and contribute about a 
5\% correction. A similar pattern of the relative impact of the 
corrections is observed if $\tb$ is varied while 
$M_{H^\pm}$ is adjusted to  keep $M_{h_1}$ constant at $40\,\mathrm{GeV}$ 
(not shown in the plot). Values of $\tb$ below 5 yield a significant
increase of the decay width.

\begin{figure}[htb]
\centering
\subfigure[]{\label{MUE1}\includegraphics[height=7.5cm]{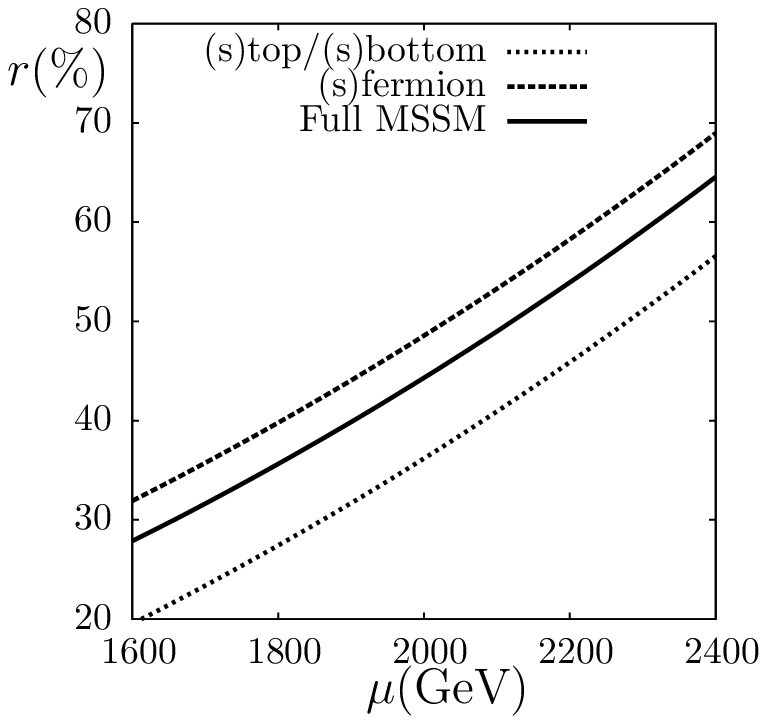}}
\subfigure[]{\label{argAtratio1}\includegraphics[height=7.5cm]{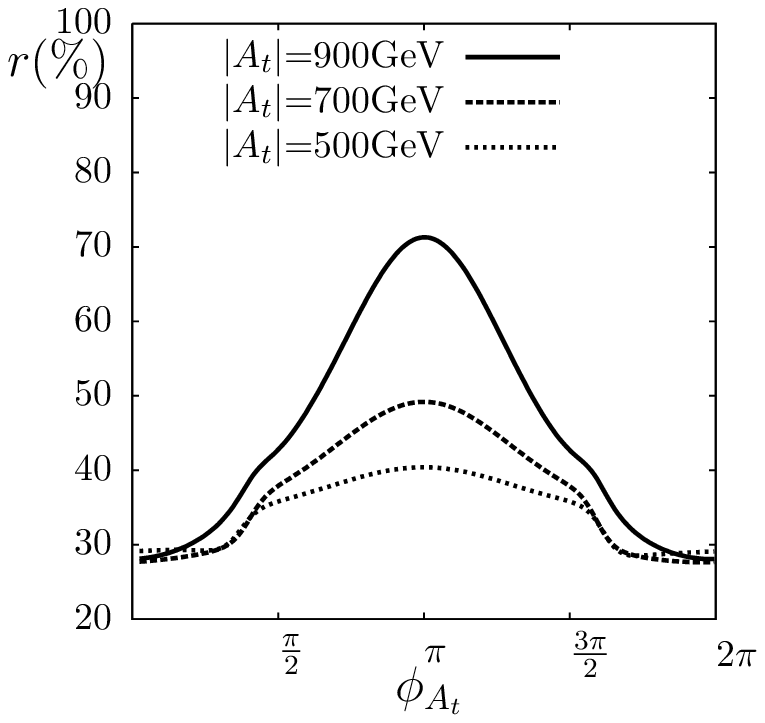}}
 \caption{\footnotesize{The ratio $r=(\Gamma_{\mathrm{Full\;Loop}}-\Gamma_{\mathrm{Improved\,Born}})/\Gamma_{\mathrm{Improved\,Born}}$ for $\tilde{\chi}^0_2\rightarrow\tilde{\chi}^0_1 h_1$
in the CPX scenario. (a) $r$ plotted against $\mu$, with various subsets of diagrams included.  $M_{H^{\pm}}$ was adjusted in order to keep $M_{h_1}=40\,\mathrm{GeV}$ constant and $\tan\beta=5.5$.  
(b) $r$ plotted against $\phi_{A_t}$.  The solid, dashed and dotted
curves correspond to $|A_t|=900,700,500\,\mathrm{GeV}$, respectively.  $M_{H^{\pm}}$ was adjusted in order to keep $M_{h_1}=45\,\mathrm{GeV}$ constant and $\tan\beta=7$.  (A Higgs mass of $M_{h_1}=40\,\mathrm{GeV}$ was not theoretically accessible for all $\phi_{A_t}$ when $|A_t|=500\,\mathrm{GeV}$.) }}
\label{MUE}
\end{figure}

Such large effects from the genuine vertex corrections are not unexpected 
in the CPX scenario (see also \citere{Williams:2007dc} for an analysis
of genuine vertex corrections to Higgs cascade decays).
  It is well known that loop corrections in the Higgs sector can be large, 
especially in this rather extreme scenario
 with large trilinear couplings and $\cp$-violating phases.  Such a large value of $\mu$ also enhances the effect
 of loop corrections in the neutralino sector.  In Figure \ref{MUE1} we
see how the effect of the genuine vertex
 corrections is further enhanced to values of 60\% or more 
if $\mu$ is increased compared to its
value in the CPX scenario of $\mu = 2\,\mathrm{TeV}$.
On the other hand, if $\mu$ is decreased one obtains correspondingly 
smaller corrections.

We also examined the effect of varying the magnitude and $\cp$-violating 
phase of the trilinear coupling, $A_t= A_b = A_{\tau}$, for the third
 generation of sfermions.  In Figure \ref{argAtratio1}, we plot $r$ as a 
function of $\phi_{A_t}$ for various values of $|A_{t}|$.   First we discuss the bold curve, where $|A_{t}|=900\,\mathrm{GeV}$.
  At $\phi_{A_t}=\pi/2$, the loop corrections show a steep dependence on the phase, $\phi_{A_t}$, emphasising the 
importance of including these phases in the calculation.  At this value,
$h_1$ has its largest $\cp$-odd content, i.e.\
 $|\mathbf{\hat{Z}}_{13}|$ is largest, while the $\cp$-even contributions 
are suppressed, giving rise to corrections of 
order $r\sim45\%$ (see Figure \ref{Mh1ratio}).
When $\phi_{A_t}=0$, the loop corrections are found to be somewhat
smaller, leading to $r\sim30\%$.  On the other hand, 
the effect of the genuine vertex corrections is maximised for 
$\phi_{A_t}=\pi$, i.e.\ $A_t=-|A_t|$. This
 corresponds to a maximum in $|\mathbf{\hat{Z}}_{11}|$, so that the lightest Higgs boson is mostly $\cp$-even.  The 
genuine vertex corrections for a $\cp$-even Higgs are larger than for a 
$\cp$-odd one, so that their effect is maximised here. Hence the
corrections in such a $\cp$-conserving scenario can even exceed the ones
in the CPX scenario. It should be noted in this context, however, that 
such a light $\cp$-even Higgs boson is of course experimentally excluded.
For smaller values of $|A_{t}|$ (dotted curves in Figure
\ref{argAtratio1}) the corrections are in general smaller, and the
variation with the phase of $A_{t}$ is less pronounced. Nevertheless,
even for $|A_t|=500\,\mathrm{GeV}$ we find $r\sim35\%$ and $r\sim40\%$ at
$\phi_{A_t} = \pi/2$ and $\phi_{A_t} = \pi$, respectively. 

\begin{figure}[htb]
 \centering
 \subfigure[]{\label{smallalphaeffMUETB10a}\includegraphics[height=7.5cm]{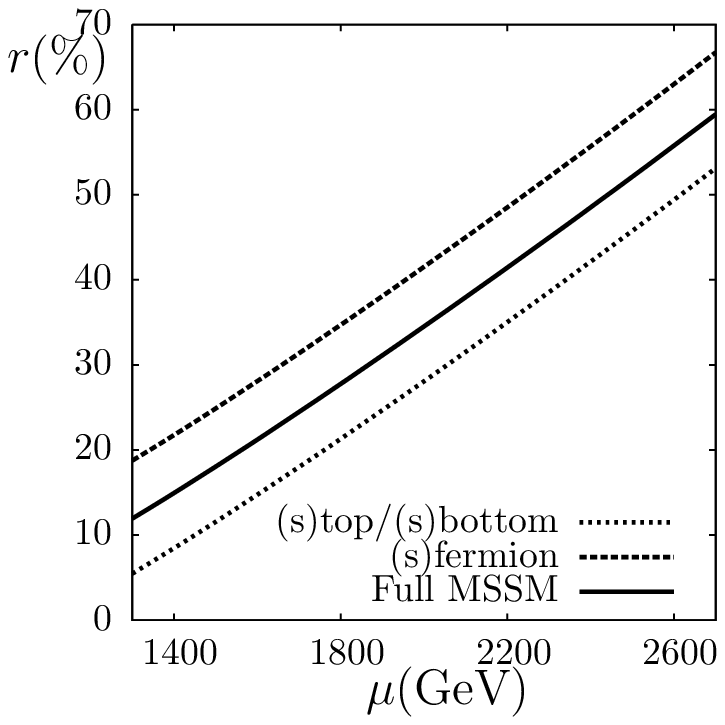}}
 \subfigure[]{\label{smallalphaargAtvaryTB}\includegraphics[height=7.5cm]{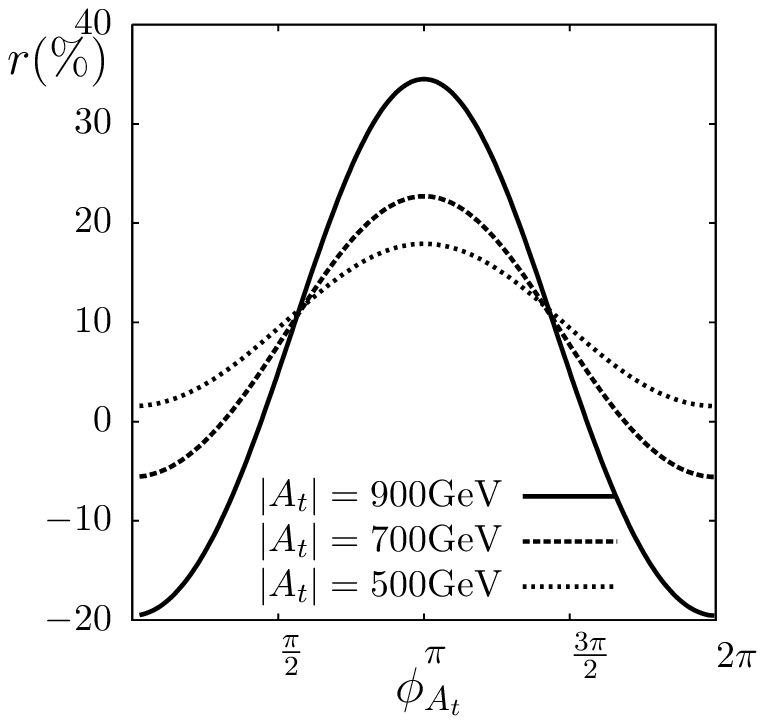}}
  \caption{\footnotesize{The ratio
$r=(\Gamma_{\mathrm{Full\;Loop}}-\Gamma_{\mathrm{Improved\,Born}})/\Gamma_{\mathrm{Improved\,Born}}$
for $\tilde{\chi}^0_2\rightarrow\tilde{\chi}^0_1 h_1$ in the small
$\alpha_{\mathrm{eff}}$ scenario.  (a) r plotted against $\mu$ with
various subsets of diagrams included.  In this plot
$M_{H^{\pm}}=220\,\mathrm{GeV}$ and $\tan\beta=10$. b) r plotted against
$\phi_{A_t}$ for three different values $\tan\beta=3.3,5,10$ and hence
$|A_t|=500,700,900\,\mathrm{GeV}$ respectively (see \refeq{sfermion}), 
with $M_{H^{\pm}}=220\,\mathrm{GeV}$. }}
 \label{smallalphaeffMUE}
 \end{figure}

We next consider the small $\alpha_{\mathrm{eff}}$ scenario.  Like the CPX scenario, this scenario has
large $\mu$ and large, negative $A_t$.  For the small
$\alpha_{\mathrm{eff}}$ scenario with $M_{H^{\pm}}=220\,\mathrm{GeV}$ and 
$\tan\beta=10$, we find genuine vertex corrections of size $r\sim35\%$.
The variation with $\mu$, shown in Figure \ref{smallalphaeffMUETB10a},
results in a pattern that is very similar to the one observed for the 
CPX scenario in \ref{MUE1}. The size of the correction scales approximately linearly
with $\mu$, and the inclusion of the full (s)fermion contributions
yields a shift of about 10\% compared to the contribution of only the
third generation (s)quarks. The non-(s)fermionic corrections to the 
genuine vertex give rise to a downward shift of about 5\%.
In Figure \ref{smallalphaargAtvaryTB} the small $\alpha_{\mathrm{eff}}$
scenario is modified by varying the phase $\phi_{A_t}$ while keeping 
$|A_t|=|X_t-\mu^{\ast}\cot\beta|$ constant. 
We find that, like the CPX scenario,
 the genuine vertex corrections have the largest effect of order
 $35\%$ at the nominal value of $\phi_{A_t}=\pi$, while the corrections are only a few percent when the phase is 
maximally CP-violating for $\phi_{A_t}=\pi/2$.  This can again be compared with Figure \ref{argAtratio1}.  Unlike
 the CPX scenario for which the corrections are minimised at $\phi_{A_t}=0,2\pi$, in the small $\alpha_{\mathrm{eff}}$
 scenario the vertex corrections exhibit another extremum here, with $r\sim-20\%$.  As for Figure \ref{MUE}(b), 
the dotted curves in Figure \ref{smallalphaargAtvaryTB} show the reduced effect of the loop corrections when $|A_t|$ is decreased.  
Here we vary $\tan\beta$ in order to produce the desired $|A_t|$ from
$X_t=-1100$GeV using \refeq{sfermion}.

In addition to the CPX and small $\alpha_{\mathrm{eff}}$ scenarios, 
we examined all of the kinematically open decay
 modes of the form $\tilde{\chi}^0_i\rightarrow\tilde{\chi}^0_j h$ for 
the SPS benchmark points~\cite{Allanach:2002nj}.
Corrections  of over $10\%$ to the 
partial decay widths were found to be common, indicating that significant 
effects are not limited to scenarios with very large values of $\mu$.  
Another scenario which we examined 
was the benchmark point LM5, studied in the CMS Technical Design Report, 
in the context of the decay 
$\tilde{\chi}^0_2\rightarrow\tilde{\chi}^0_1 h$~\cite{Ball:2007zza}.
  For this scenario we found the corrections to the partial decay width to be around $5\%$.
  However, due to the large branching ratio of $85\%$ for the process, these corrections translated
 into an effect of less than a percent on the branching ratio.  

\section{Phenomenology at the LHC}
\label{sectionpheno}
\subsection{Numerical Results for Branching Ratio}
In the previous section, we found that the genuine vertex corrections to the partial decay width
 $\Gamma(\tilde{\chi}^0_{2}\rightarrow\tilde{\chi}^0_1 h_{1})$ were of order $45\%$ in the CPX
scenario.  For phenomenology at the LHC it is important to consider, in addition to the decay widths, also the branching ratios 
of neutralinos.  
In this section, we compute the branching ratios 
of $\tilde{\chi}^0_{2}$, incorporating our loop-corrected decay widths for $\tilde{\chi}^0_{2}\rightarrow\tilde{\chi}^0_1 h_{1,2,3}$.
  As well as calculating the genuine vertex corrections to $\tilde{\chi}^0_i\rightarrow \tilde{\chi}^0_j h_a$, we have also
 calculated the genuine vertex corrections to $\tilde{\chi}^0_i\rightarrow \tilde{\chi}^0_j Z$, using a similar procedure
 to that detailed in the previous sections for the Higgs vertex.  We incorporate these into the branching ratio calculation too.
In the CPX scenario, depending on its mass, $\tilde{\chi}^0_2$ can decay via the following decay modes:
\begin{eqnarray}
\tilde{\chi}^0_2&\rightarrow& \tilde{\chi}^0_1 h_1,\; \tilde{\chi}^0_1 h_2,\; \tilde{\chi}^0_1 h_3,\; \tilde{\chi}^0_1 Z,\;
\tilde{\chi}^0_1 f \bar{f},\;  \tilde{f}_{1,2} \bar{f},\; \tilde{\bar{f}}_{1,2} f.
\end{eqnarray}
Where kinematically possible, we calculate the decays $\tilde{\chi}^0_2\rightarrow \tilde{\chi}^0_1 h_a$, which produce on-shell
 neutral Higgs bosons, as two-body decays, including the genuine vertex corrections as detailed in the previous sections.  
Where kinematically possible, we also calculate the decay $\tilde{\chi}^0_2\rightarrow\tilde{\chi}^0_1 Z$ into an on-shell Z  boson
 as a two-body decay, including the equivalent genuine vertex corrections as mentioned above. 
Finally, we calculate the 3-body decay $\tilde{\chi}^0_2\rightarrow\tilde{\chi}^0_1 f \bar{f}$.  For this, we include, firstly, the
 diagrams where an off-shell Higgs boson is exchanged (i.e.\ where some or all of $h_1,h_2,h_3$ are too heavy to be
 produced on-shell).  For these diagrams we use the unitary $\mathbf{\hat{U}}$ matrix elements and masses from 
\texttt{FeynHiggs} to construct effective couplings (see Ref. \cite{Frank:2006yh}) which take into account the two-loop Higgs
 propagator-type corrections.  Secondly, in the three-body decay, where the kinematics do not permit an on-shell Z boson, we include
 the diagram where a Z boson is exchanged, along with the diagram where the would-be Goldstone boson, $G$, is
 exchanged (in this way a proper cancellation of the gauge dependence is ensured).  Thirdly, we include in the three-body decay the diagrams where a sfermion is
 exchanged.  As the neutralino mass approaches the scale of the sfermion masses, the 
possibility of on-shell production of sfermions
 arises, which 
subsequently decay into $\tilde{\chi}^0_1$.  To take account of this threshold region, we include a finite width for each
 sfermion, calculated from its self-energy.  All self-energies and two- and three-body partial decay widths were calculated with the
 help of \texttt{FeynArts} and \texttt{FormCalc}.

Throughout most of the CPX parameter space, $\tilde{\chi}^0_2\rightarrow\tilde{\chi}^0_1 Z$ can proceed
 as a two-body decay.  For the CPX scenario, we find that the genuine vertex corrections to this decay width can be of order $30\%$.  However, the amplitude is suppressed by several orders of magnitude in the CPX scenario, since the Z boson
 only couples to the higgsino component of each of the neutralinos, while the large value of $\mu$ renders 
$\tilde{\chi}^0_{1}$ and $\tilde{\chi}^0_{2}$  mostly bino and wino, respectively.  

\begin{figure}[htb]
\centering
\subfigure[]{\label{M2bBR}\includegraphics[height=7.5cm]{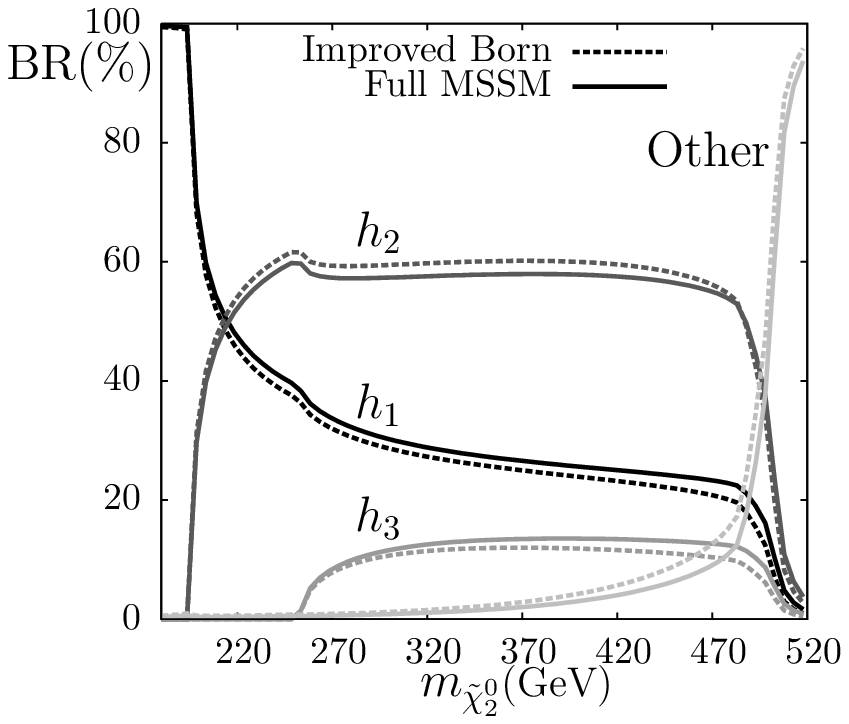}}
\subfigure[]{\label{Mh1BR}\includegraphics[height=7.5cm]{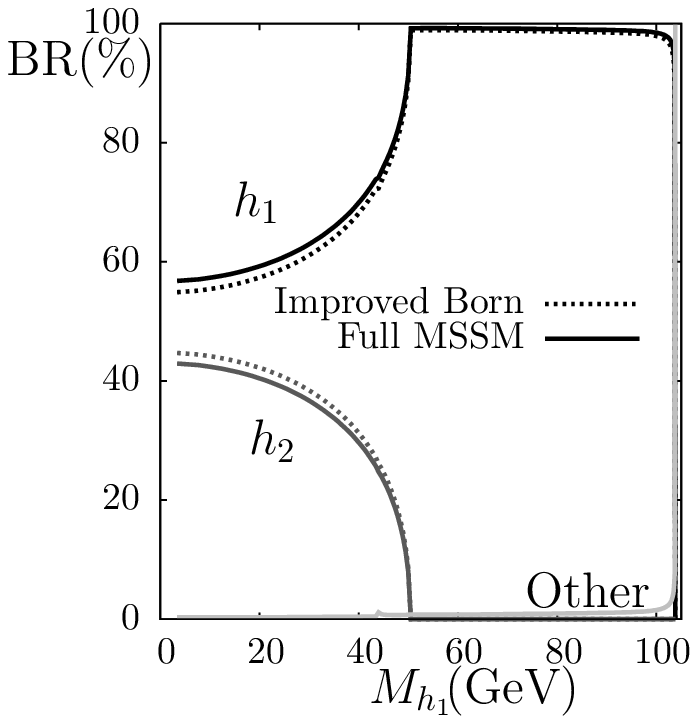}}
\caption{\footnotesize{Branching ratio for each of $\tilde{\chi}^0_2\rightarrow \tilde{\chi}^0_1 h_{1,2,3}$ and for the other decay modes, $\tilde{\chi}^0_2\rightarrow \tilde{\chi}^0_1 Z$ and $\tilde{\chi}^0_2\rightarrow \tilde{\chi}^0_1 f\bar{f}$ (labelled ``Other"); (a) shown as a function of $m_{\tilde{\chi}^0_2}$, for $M_{h_1}=40\,\mathrm{GeV}$ and $\tan\beta=5.5$ ($M_2$ was varied as input to produce the change in $m_{\tilde{\chi}^0_2}$); and, (b) shown as a function of $M_{h_1}$ for $\tan\beta=5.5$ and $M_2=200\,\mathrm{GeV}$ ($M^{H^{\pm}}$ was varied as input).  In both plots we show the Improved Born approximation as a dashed line and the full MSSM result as a solid line.}}
\label{allBR}
\end{figure}
The resulting branching ratios of $\tilde{\chi}^0_2$ in the CPX scenario are plotted as a function 
of the neutralino mass, $m_{\tilde{\chi}^0_{2}}$, in Figure \ref{allBR}a, with $\tan\beta=5.5$ and $M_{h_1}=40\,\mathrm{GeV}$. 
Both the Improved Born and full MSSM vertex-corrected results are shown.  We see that 
for $m_{\tilde{\chi}^0_{2}}\lsim190\,\mathrm{GeV}$, $\mathrm{BR}(\tilde{\chi}^0_2\rightarrow\tilde{\chi}^0_1 h_1)\approx 100\%$, 
and therefore the loop corrections to the $\tilde{\chi}^0_2\rightarrow \tilde{\chi}^0_1 h_{1}$ partial width have negligible effect.  As one increases $m_{\tilde{\chi}^0_{2}}$ from $190$ to 
$470\,\mathrm{GeV}$, the on-shell decays
$\tilde{\chi}^0_2\rightarrow\tilde{\chi}^0_1 h_2$ and $\tilde{\chi}^0_2\rightarrow\tilde{\chi}^0_1 h_3$
 become kinematically allowed.  This causes $\mathrm{BR}(\tilde{\chi}^0_2\rightarrow\tilde{\chi}^0_1 h_1)$ to vary from 
$100\%$ to around $25\%$.  In this region, the three competing decay modes into Higgs bosons all receive large 
vertex corrections of order $50\%$.  However, since these vertex corrections have similar structure, their 
effects tend to cancel each other out, producing an effect of only a few percent on the branching ratios.
Thus, the Improved Born approximation works well in this region.  

The effect of vertex corrections on the branching ratio will be more significant in regions of parameter space where there is 
another competing decay mode of $\tilde{\chi}^0_2$ which does not have loop corrections of a similar 
structure to  $\tilde{\chi}^0_2\rightarrow \tilde{\chi}^0_1 h_1$.  
In the CPX scenario, this competition will never be provided by the highly suppressed decay into a Z boson.  
However, for $m_ {\tilde{\chi}^0_2}$ large enough, decays via sfermions
 become important.  While the Higgs bosons require both a non-zero gaugino and higgsino component to couple to neutralinos,
 sfermions couple only to the gaugino part.  Thus, if sfermion decays can
 proceed on-shell, they will, in this scenario, dominate over
 the Higgs decay modes, rendering $\mathrm{BR}(\tilde{\chi}^0_2\rightarrow\tilde{\chi}^0_1 h_1)\approx 0$.  Before this,
 there will be a threshold region, which can be seen in Figure \ref{M2bBR} for $450\lsim m_{\tilde{\chi}^0_2}\lsim 520\,\mathrm{GeV}$.  Within
 this region, the existence of competing decay modes means that the genuine vertex corrections are very important.  
The maximum effect occurs near $m_{\tilde{\chi}^0_2}\sim500$GeV, where the positive vertex corrections to the Higgs decay
 widths result in a reduction of the branching ratio $\mathrm{BR}(\tilde{\chi}^0_2\rightarrow\tilde{\chi}^0_1 f \bar{f})$ 
of more than $10\%$ compared to its Improved Born value. 

In Figure \ref{Mh1BR}, we show the branching ratios of $\tilde{\chi}^0_2$ as a function of $M_{h_1}$, to be compared 
with Figure \ref{Mh111}.  Here $M_2=200\,\mathrm{GeV}$ and $\tan\beta=5.5$, so only decays into $\tilde{\chi}^0_1 h_1$, $\tilde{\chi}^0_1 h_2$, $\tilde{\chi}^0_1 Z$ and $\tilde{\chi}^0_1f\bar{f}$ (the latter two labelled ``Other") are kinematically open.  For $M_{h_1}\gsim50\mathrm{GeV}$, the second lightest Higgs boson is too heavy 
to be produced on-shell and so $\mathrm{BR}(\tilde{\chi}^0_2\rightarrow\tilde{\chi}^0_1 h_1)$ is close to $100\%$.  In the 
CPX hole, with $M_{h_1}\approx40$GeV, we find $\mathrm{BR}(\widetilde{\chi}^0_2\rightarrow\widetilde{\chi}^0_1 h_1)\approx79\%$, an increase of around $3\%$ compared to the Improved Born value. 

\begin{figure}[htb]
\centering
\subfigure[]{\label{testTB}\includegraphics[height=7.5cm]{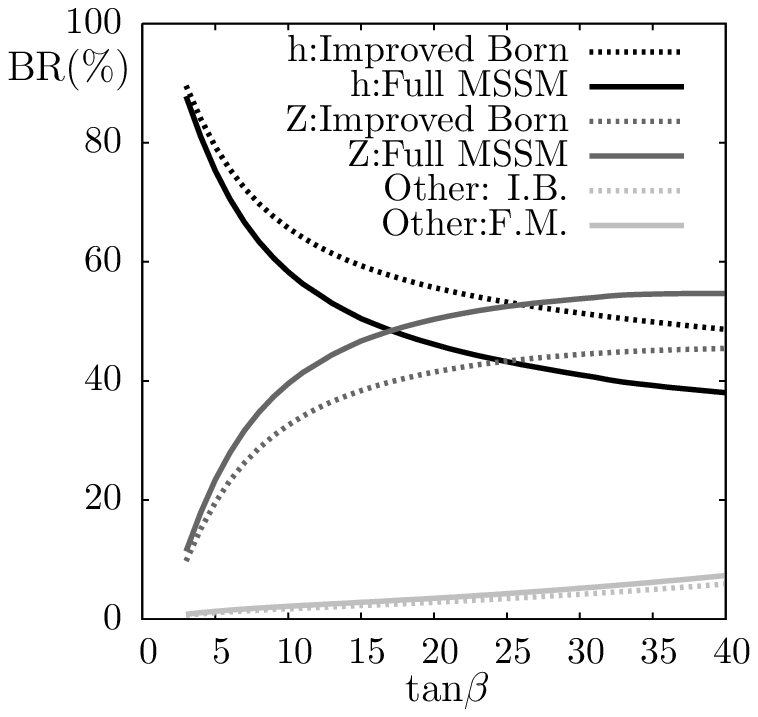}}
\subfigure[]{\label{testAt}\includegraphics[height=7.5cm]{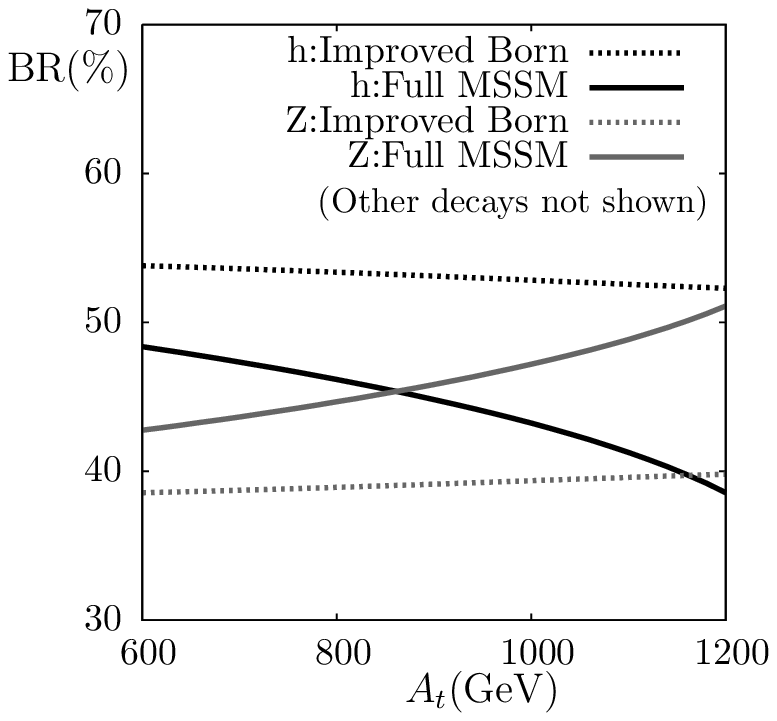}}
\caption{\footnotesize{Branching ratios for each of $\tilde{\chi}^0_2\rightarrow \tilde{\chi}^0_1 h$, $\tilde{\chi}^0_2\rightarrow \tilde{\chi}^0_1 Z$ and $\tilde{\chi}^0_2\rightarrow \tilde{\chi}^0_1 f\bar{f}$ (labelled ``Other'') in the $\cp$-conserving ``light $\tilde{\chi}^0_1$ scenario": (a) shown as a function of $\tan\beta$, for fixed $A_f=1\,\mathrm{TeV}$; and, (b) shown as a function of $A_t$, for fixed $\tan\beta=20$. We show the Improved Born approximation (I.B.) as the dotted line, and the full MSSM result (F.M.) as the solid line.}}
\label{lightchi}
\end{figure}

Thus, although we found large loop corrections to the partial decay widths of $\tilde{\chi}^0_2\rightarrow \tilde{\chi}^0_1 h_{1,2,3}$
 in the CPX scenario, the effects on the branching ratios turn out to be significantly smaller, because the decays are 
not competing with other modes and so the large genuine vertex corrections cancel each other out.  
This will also be the case for the small $\alpha_{\mathrm{eff}}$ scenario,
 in which the Z decay mode is also suppressed and the sfermions are heavy.  However, this situation is not generic, and large vertex corrections
 can affect the branching ratios if there are other competing decay modes with vertex corrections of a different structure.
   In non-gaugino-like scenarios,
 without a large hierarchy between $M_2$ and $\mu$, the decays into Higgs bosons are more likely to compete 
with the decays into Z bosons and sfermions.
  For example, in Figure \ref{lightchi} we show the Improved Born and full MSSM branching ratios for the  
``light $\tilde{\chi}^0_1$ scenario" of Eqn. \ref{eqnlightchi1}.  Here we can have 
$\mathrm{BR}(\tilde{\chi}^0_2\rightarrow \tilde{\chi}^0_1 h)\sim \mathrm{BR}(\tilde{\chi}^0_2\rightarrow \tilde{\chi}^0_1 Z)$. 
 We computed genuine vertex corrections to both $\Gamma(\tilde{\chi}^0_2\rightarrow \tilde{\chi}^0_1 h)$ and 
$\Gamma(\tilde{\chi}^0_2\rightarrow \tilde{\chi}^0_1 Z)$, and found the former (latter) to be negative and of order $20\%$ ($3\%$) and $35\%$ ($2\%$)
for $A_t=500\,\mathrm{GeV}$ and $A_t=1200\,\mathrm{GeV}$, respectively. The corrections are further enhanced at large values of $\tan\beta$.  
In this scenario, the branching ratio for $\tilde{\chi}^0_2\rightarrow \tilde{\chi}^0_1 h$ happens to be near $50\%$.  
Thus, the effect of the vertex corrections on the branching ratios is maximised in this case.  The plots in Figure \ref{lightchi} show 
corrections to the branching ratio of more than $10\%$.

\subsection{Prospects for the ``CPX Hole"}
In the previous section, we found that $\tilde{\chi}^0_2\rightarrow\tilde{\chi}^0_1 h_1$ has a large branching ratio, $\mathrm{BR}(\widetilde{\chi}^0_2\rightarrow\widetilde{\chi}^0_1 h_1)\sim79\%$, for
 the ``CPX hole", i.e.\ in the region where a light Higgs is unexcluded by present data. We now investigate whether Higgs production in neutralino decays at the LHC could help to cover this parameter region.
Consider the SUSY cascade decay chain starting with a gluino;
\begin{equation}
 \widetilde{g} \rightarrow \widetilde{f} \bar{f}\rightarrow \widetilde{\chi}^0_2 f \bar{f} \rightarrow \widetilde{\chi}^0_1 f \bar{f} h_i  \rightarrow \widetilde{\chi}^0_1 f \bar{f} b \bar{b} (\tau^+ \tau^-). 
\end{equation}
Coloured sparticles like gluinos are expected to be produced in large numbers at the LHC provided they are light enough, 
(see eg. Ref.~\cite{Weiglein:2004hn} for 
detailed analyses of SUSY cascade decays).  
  These gluinos will decay into lighter coloured sparticles, namely the squarks with masses around 
$500\,\mathrm{GeV}$ (see Table \ref{tab:masses}).  For most squarks, the only way to conserve R-parity will be
 to decay into $\tilde{\chi}^0_2$, $\tilde{\chi}^{\pm}_1$ and $\tilde{\chi}^0_1$.  As shown in the previous section, $79\%$ of the produced $\tilde{\chi}^0_2$ will decay into $h_1$ in this scenario for a Higgs mass of $40\,\mathrm{GeV}$.  The light Higgs boson then  
decays mostly into $b\bar{b}$ ($91\%$), and also $\tau^+\tau^-$. 

Branching ratios for all parts of the decay chain, except the decays involving Higgs bosons, were computed at tree level using \texttt{FeynArts} and \texttt{FormCalc}.  For decays involving Higgs bosons, such as $\tilde{t}_2\rightarrow\tilde{t}_1 h_a$, we use an Improved Born approximation, as in Eq.~\ref{eqnib}.
We computed the branching ratio for $\widetilde{g}\rightarrow\tilde{q}_{1,2} q$ for each of $q=u,d,c,s,t,b$.  We found that
 $\tilde{b}_1$, $\tilde{u}_1$, $\tilde{d}_2$, $\tilde{c}_2$, $\tilde{s}_1$, $\tilde{s}_2$
 all have substantial branching ratios to decay into $\tilde{\chi}^0_2$.  Summing over the 
various decay modes we found that $17\%$ of all gluinos produced decay via a squark into 
$\widetilde{\chi}^0_2$.

Combining $\mathrm{BR}(\widetilde{\chi}^0_2\rightarrow\widetilde{\chi}^0_1 h_1)\sim79\%$ with 
$\mathrm{BR}(\widetilde{g}\rightarrow\widetilde{\chi}^0_2 q \bar{q})\sim17\%$, we estimate that around 
$13\%$ of the gluinos produced in this scenario will decay into $h_1$.  
Thus, SUSY cascade decays where a light Higgs is produced in the decay 
of the second-lightest neutralino appear to be a promising possibility
to cover this problematic parameter region where standard search
channels may only have small sensitivities. Detailed experimental
analyses would be needed to determine whether it is indeed possible in
such a case to extract a Higgs signal from the SM and SUSY backgrounds.%
\footnote{It should be noted in this context that the CMS collaboration
has performed a full detector simulation and event reconstruction for
the production of a Higgs boson at the end of a cascade of
supersymmetric particles starting with squarks and
gluinos~\cite{Ball:2007zza}. These results, obtained for the benchmark
point LM5, cannot be directly translated to the case of the CPX
scenario, since in the case of LM5 the Higgs boson is much heavier,
$m_{h}\sim 115\gev$, than in the region of the CPX scenario that we are
considering here. The $b$ jets resulting from the Higgs decay in the CPX
scenario are therefore softer than for LM5, so that cuts on the
energy of the jets will be less efficient to suppress the QCD
background.}

\section{Conclusions}

We have obtained complete one-loop results for the class of processes 
involving the decay of a neutralino into a 
neutral Higgs boson plus a lighter neutralino, 
$\tilde{\chi}^0_i\rightarrow\tilde{\chi}^0_j h_a$. 
The genuine vertex contributions to the neutralino decay amplitudes
have been combined with state-of-the-art two-loop propagator-type 
corrections for the outgoing Higgs boson.
Our results take into account all sectors of the MSSM and include the 
full phase dependence of the $\cp$-violating parameters $A_f$ and $M_3$.  

For the renormalisation in the chargino--neutralino sector, 
we have worked out an on-shell scheme which properly takes into account 
imaginary parts arising from complex parameters and from absorptive parts
of loop integrals. In this scheme in- and
outgoing fermions receive different field renormalisation constants.
Since we have concentrated in this paper in particular on the CPX
benchmark scenario, where only the parameter $M_3$ and the trilinear
couplings of the third generation fermions are complex, we have not
specified the renormalisation of the phases of the parameters appearing
in the neutralino and chargino mass matrices. This issue will be
addressed in a forthcoming publication.

In the CPX scenario we find corrections to the partial decay width for
 $\widetilde{\chi}^0_2\rightarrow\widetilde{\chi}^0_1 h_1$ of about
 $45\%$ relative to the Improved Born
 approximation.  These corrections, which characterise the impact of the
genuine vertex contributions, scale almost linearly with the higgsino
parameter, $\mu$. We also found a strong dependence of the size of the 
genuine vertex corrections on the absolute value, $|A_t|$, and the
$\cp$-violating phase, $\phi_{A_t}$, of the third generation sfermion
trilinear coupling. The corrections turn out to be even larger for the 
($\cp$-conserving) case where $\phi_{A_t}=\pi$ compared to the case of
the maximally $\cp$-violating phase of $\phi_{A_t}=\pi/2$.

We have also investigated $\cp$-conserving scenarios,
in particular the small 
$\alpha_{\mathrm{eff}}$ scenario. Similar to the CPX scenario, the small
$\alpha_{\mathrm{eff}}$ scenario is characterised by large values of
$\mu$ and $|A_t|$, and we found large corrections to the partial decay
width of about $35\%$. In both 
the CPX and small $\alpha_{\mathrm{eff}}$ scenarios, the predominantly
gaugino-like character of the two light neutralinos
 reduces the effect of the large vertex corrections on the branching ratios 
down to a few percent.  
However, we also applied our results to the ``light $\tilde{\chi}^0_1$''
scenario and found 
corrections to the branching ratio of more than $10\%$.  We also investigated a number of other 
$\cp$-conserving scenarios and found non-negligible corrections.  
Our results can also be applied to the process 
$h_a\rightarrow \tilde{\chi}^0_i \tilde{\chi}^0_j$.

Based on our precise predictions for 
$\Gamma(\widetilde{\chi}^0_2\rightarrow\widetilde{\chi}^0_1 h_1)$ 
and the corresponding branching ratio, 
we have investigated the prospects of SUSY cascade decays, where a Higgs
is produced in the decay of a neutralino, for covering the parameter region of
the CPX scenario in which a light Higgs boson is unexcluded.
We find that around $13\%$ of all gluinos produced at the LHC in the 
CPX scenario will decay via the second lightest neutralino into the 
lightest Higgs boson. Thus, Higgs production in neutralino decays
looks promising as a search channel for such a light Higgs, while
standard search channels may have small sensitivities in this parameter
region. The results obtained in this paper will be provided as a public
tool with the aim of facilitating further experimental studies of this
potentially interesting channel.

\section*{Acknowledgements}
We thank Thomas Hahn, Sven Heinemeyer, Olaf Kittel, Sophy Palmer, Krzysztof Rolbiecki, 
Christian Schappacher, Markus Schumacher
and Karina Williams for numerous helpful discussions.  
This work has been supported
in part by the European Community's Marie-Curie Research
Training Network under contract MRTN-CT-2006-035505
`Tools and Precision Calculations for Physics Discoveries at Colliders'
(HEPTOOLS) and MRTN-CT-2006-035657
`Understanding the Electroweak Symmetry
Breaking and the Origin of Mass using the First Data of ATLAS'
(ARTEMIS).  AF acknowledges support by a Commonwealth Scholarship and a Durham University Postgraduate Teaching Fellowship.

\providecommand{\href}[2]{#2}\begingroup\raggedright\endgroup

\end{document}